\documentclass[hyper]{JHEP3}

\input{epsf}
\usepackage{amsfonts}
\usepackage{amsbsy}
\usepackage{amsmath}
\usepackage{amssymb}
\usepackage{epsfig}
\usepackage{cite}
\usepackage{epstopdf}

\newdimen\tableauside\tableauside=1.0ex
\newdimen\tableaurule\tableaurule=0.4pt
\newdimen\tableaustep
\def\phantomhrule#1{\hbox{\vbox to0pt{\hrule height\tableaurule width#1\vss}}}
\def\phantomvrule#1{\vbox{\hbox to0pt{\vrule width\tableaurule height#1\hss}}}
\def\sqr{\vbox{%
  \phantomhrule\tableaustep
  \hbox{\phantomvrule\tableaustep\kern\tableaustep\phantomvrule\tableaustep}%
  \hbox{\vbox{\phantomhrule\tableauside}\kern-\tableaurule}}}
\def\squares#1{\hbox{\count0=#1\noindent\loop\sqr
  \advance\count0 by-1 \ifnum\count0>0\repeat}}
\def\tableau#1{\vcenter{\offinterlineskip
  \tableaustep=\tableauside\advance\tableaustep by-\tableaurule
  \kern\normallineskip\hbox
    {\kern\normallineskip\vbox
      {\gettableau#1 0 }%
     \kern\normallineskip\kern\tableaurule}%
  \kern\normallineskip\kern\tableaurule}}
\def\gettableau#1 {\ifnum#1=0\let\next=\null\else
  \squares{#1}\let\next=\gettableau\fi\next}

\title{Orientifolds and the Refined Topological String}

\author{Mina Aganagic$^1$ and Kevin Schaeffer$^1$\\
\\
${}^1$ Center for Theoretical Physics,
University of California, Berkeley.}

\abstract{We study refined topological string theory in the presence of orientifolds by counting second-quantized BPS states in M-theory.  This leads us to propose a new integrality condition for both refined and unrefined topological strings when orientifolds are present.  We define the $SO(2N)$ refined Chern-Simons theory which computes refined open string amplitudes for branes wrapping Seifert three-manifolds.  We use the $SO(2N)$ refined Chern-Simons theory to compute new invariants of torus knots that generalize the Kauffman polynomials.  At large N, the $SO(2N)$ refined Chern-Simons theory on $S^{3}$ is dual to refined topological strings on an orientifold of the resolved conifold, generalizing the Gopakumar-Sinha-Vafa duality.  Finally, we use the $(2,0)$ theory to define and solve refined Chern-Simons theory for all ADE gauge groups.}

\begin{document}

\bibliographystyle{utphys}
\section{Introduction}

Recently in \cite{CSRefined}, a refinement of \(SU(N)\) Chern-Simons theory on Seifert manifolds was constructed by studying certain M-theory backgrounds.  There it was also shown that in the large N limit, the partition function of refined Chern-Simons theory on \(S^{3}\) is equal to the partition function of the refined topological string on the resolved conifold, thus providing a refinement of the celebrated Gopakumar-Vafa duality \cite{Gopakumar:1998ki}.

It was also shown that the refined Chern-Simons theory can be used to explicitly compute new two-variable polynomials associated to torus knots.  These invariants generalize the one-variable quantum \(SU(N)\) knot invariants.  Further, the authors of \cite{CSRefined} discovered that under appropriate changes of variables, these new polynomials could be used to determine the superpolynomial \cite{Dunfield:2005si} of certain torus knots.  The superpolynomial is the Poincare polynomial of a knot homology theory categorifying the HOMFLY polynomial.  It encodes the large \(N\) behavior of Khovanov-Rozansky knot invariants.

In analogy with the unrefined case, refined Chern-Simons on a three-manifold \(M\) is equivalent to open refined string theory on the Calabi-Yau, \(T^{*}M\).  Thus refined Chern-Simons theory is part of the broader program of understanding the refinement of topological strings in both the closed and open settings.  In the absence of a worldsheet definition, this refinement is best understood in terms of counting BPS states in M-theory.

In this paper, we analyze refined topological string theory when there is an orientifold acting on the spacetime.  Again, we find that refinement is most naturally understood by studying an index that counts second-quantized BPS contributions in M-theory.  This analysis leads us to propose new integrality structures for orientifolds of both the refined string partition function, and the unrefined string studied previously in \cite{SinhaVafa, Acharya:2002ag, Bouchard:2004iu, Bouchard:2004ri, Walcher:2007qp, Krefl:2009md, Krefl:2009mw, Marino:2009mw}.

Our analysis also sheds light on the conjecture \cite{Dijkgraaf:2009pc, Krefl:2010fm} that the refined topological string at \(\beta=\frac{1}{2}, 2\) is equal to the ordinary topological string in the presence of an \(SO(N)/Sp(N)\) orientifold.  From the second-quantized M-theory perspective, it is clear that in the noncompact part of spacetime, the trace that computes the refined topological A-model at \(\beta=\frac{1}{2}, 2\) is the same as the unrefined trace when an orientifold acts on two of the spacetime directions.  However, on the internal Calabi-Yau, \(X\), the two computations are different -- in the refined case, there is an internal \(U(1)_{R}\) rotation and in the unrefined case, there is an anti-holomorphic involution acting on \(X\).  

For simple geometries, such as the Dijkgraaf-Vafa models that engineer \(SO(N)/Sp(N)\) gauge theory with matter in the symmetric or antisymmetric representation, we expect there will be no significant difference between the presence or absence of the involution and the conjectured correspondence will hold \cite{Dijkgraaf:2002fc, Aganagic:2003xq, Cachazo:2003kx, Landsteiner:2003ph, Intriligator:2003xs}.  However, from this line of reasoning, we expect that the correspondence will not hold more generally unless we drop the involution acting on the internal Calabi-Yau.

Having understood the general structure of refined topological strings in the presence of orientifolds, we turn to the \(SO(2N)\) refined Chern-Simons theory that arises from wrapping branes on an orientifold plane. We explain how to solve the theory, and study the invariants that come from the expectation value of torus knots.  We also study the large N limit of the theory and show that it is consistent with the expected form of refined closed strings propagating on an orientifold of the resoved conifold.  This gives a new refinement of the \(SO(N)\) Chern-Simons geometric transition studied in \cite{SinhaVafa}.

The organization of this paper is as follows.  In section \ref{sec:review}, we review the construction of open and closed refined topological string theory from M-theory.  We also review the connection to instanton counting, vortex counting, and knot theory.  In section \ref{sec:ofolds}, we introduce orientifolds and explain how refinement can be extended to unoriented strings.  From this analysis we propose a new integrality condition in Section \ref{sec:integrality} for both unrefined and refined topological strings on orientifolds.  

In section \ref{sec:refinedcs}, we review the definition of refined Chern-Simons theory and its connection with refined open string theory.  We also explain the straightforward generalization of refined Chern-Simons theory to all ADE gauge groups. In section \ref{sec:knots} we clarify the connection between refined Chern-Simons theory and knot homology, and use the \(SO(2N)\) refined Chern-Simons theory to compute invariants of torus knots that generalize the Kauffman polynomials. We obtain new knot invariants associated to the \(SO(2N)\) gauge group in the fundamental and spinor representations. Unlike the \(SU(N)\) case, we find that these polynomials cannot generally be related to the Kauffman superpolynomial \cite{Gukov:2005qp} of \(SO(N)\) knot homology by a change of variables. This is not unexpected as, by construction, refined Chern-Simons theory computes an index, and not a Poincare polynomial.

Finally, in section \ref{sec:largen} we study the large N limit of the \(SO(N)\) refined Chern-Simons theories on \(S^{3}\).  The result is naturally interpreted as the partition function of refined closed strings on an orientifold of \(\mathcal{O}(-1)\oplus\mathcal{O}(-1) \to \mathbb{P}^{1}\).

In Appendix \ref{sec:twist} we give a detailed description of the topologically twisted \((2,0)\) theory on Seifert manifolds.  Some useful facts about Macdonald polynomials are reviewed in Appendix \ref{app:mac}, and specific results about \(SO(2N)\) and its Macdonald polynomials are given in Appendix \ref{app:so2n}.  Finally, in Appendix \ref{sec:index} we review the refined indices in five and three dimensions that compute the refined topological string.  

\section{M-Theory and Refined Topological Strings \label{sec:review}}
In this paper we will mainly focus on a one-parameter deformation of topological string theory that is known as the refined topological string.  Before discussing refinement, we review some useful facts about unrefined topological string theory.  Recall that the (A-model) closed topological string localizes on holomorphic maps from Riemann surfaces into a Calabi-Yau, \(X\), and is only sensitive to the Kahler structure of \(X\).   We can also introduce branes wrapping Lagrangian three-cycles, \(L \subset X\), so that the open topological string localizes on holomorphic two-chains with boundary on \(L\).  Although topological string theory was originally defined from this worldsheet point of view, it was later realized that both open and closed topological strings naturally compute certain physical indices in M-theory \cite{Gopakumar:1998ii, Gopakumar:1998jq, Hollowood:2003cv, Dijkgraaf:2006um, Ooguri:1999bv, Aganagic:2009cg, Cecotti:2010fi, Cheng:2010yw}.  In many ways, this modern viewpoint is advantageous since it reveals an integrality structure that is hidden on the worldsheet.

Although there is not yet a worldsheet definition of the refined topological string, it has a very natural definition in M-theory.  Refining the topological string simply corresponds to computing a more general trace, which is a five-dimensional analogue of the refined spin character of \cite{Gaiotto:2010be}.

In this section we review the definition of the refined topological string in both the closed and open cases.  Further, we explain how this is connected to K-theoretic instanton counting and knot homology, and emphasize the integrality properties that arise from M-theory.  This will lay the foundation for introducing Calabi-Yau orientifolds in the next section. 

In the pioneering work of \cite{Gopakumar:1998ii, Gopakumar:1998jq}, the topological A-model was reinterpreted in terms of integer Gopakumar-Vafa invariants that count BPS M2 branes.  The connection was made by starting with IIA physical string theory on the geometry, \(X \times \mathbb{R}^{3,1}\) and considering the low-energy theory in the four spacetime dimensions.  It is known that the topological A-model computes terms in the low energy effective action of the form,
\begin{equation}
\int d^{4}x \int d^{4}\theta \; F_{g}(t_{i})\big(\mathcal{W}^{2}\big)^{g}
\end{equation}
where \( F_{g}(t_{i})\) is the genus \(g\) free energy of the topological A-model, the \(t_{i}\) are the vector multiplets whose lowest components parametrize the Kahler moduli space of \(X\), and \(\mathcal{W}\) is the \(\mathcal{N}=2\) Weyl multiplet.  Expanding this out in components gives,
\begin{equation}
\int d^{4}x F_{g}(t_{i})\big(\lambda^{2}\big)^{g-1}R_{+}^{2} + \ldots
\end{equation}
where \(\lambda\) is the self-dual graviphoton field strength and \(R_{+}\) is the self-dual part of the Riemann tensor.  The crucial observation of Gopakumar and Vafa was that by turning on a background graviphoton field-strength, \(\lambda = g_{s}dx_{1}\wedge dx_{2} + g_{s}dx_{3}\wedge dx_{4}\), these terms could be reproduced by integrating out massive BPS matter.  This charged matter comes precisely from D2-D0 bound states wrapping two-cycles in \(X\).  It is natural to go to large coupling so that IIA becomes M-theory, and the D2-D0 states lift to M2 branes with momentum around the M-theory circle.  Integrating out these states by a Schwinger-type calculation gives a new way of writing the topological string free energy,
\begin{equation}
F_{top} = \sum_{\beta, s_{L}, s_{R}}\sum_{d=1}^{\infty}\sum_{j_{L}=-s_{L}}^{s_{L}}\sum_{j_{R}=-s_{R}}^{s_{R}}\frac{1}{d}\frac{(-1)^{2s_{L}+2s_{R}}N^{s_{L},s_{R}}_{\beta}q^{2 d j_{L}}Q^{\beta d}}{(q^{d/2}-q^{-d/2})^{2}}
\end{equation}
where \(q=e^{i g_{s}}\) and \(N^{s_{L},s_{R}}_{\beta}\) counts M2 branes wrapping the cycle \(\beta \in H_{2}(X,\mathbb{Z})\) with intrinsic spin, \((s_{L},s_{R})\).  It important to note that the Gopakumar-Vafa invariants \(N^{j_{L},j_{R}}_{\beta}\) are in fact \emph{integers}, since they count BPS states.  Thus by integrating out \(M2\)-branes, this gives a new integrality structure underlying topological string theory.  

Also note that \(F_{top}\) only depends on the combinations, 
\begin{equation}
N^{s_{L}}_{\beta} \equiv \sum_{s_{R}}(-1)^{2s_{R}}(2s_{R}+1)N^{s_{L}, s_{R}}_{\beta}
\end{equation}
so these are usually defined as the Gopakumar-Vafa invariants.  The starting point for refining the topological string is the desire to define a more general theory that keeps track of the full spin information, and not just the \(s_{L}\) spin content of \(M2\) branes.  This can be done by turning on a non-self-dual graviphoton background, \(\lambda = \epsilon_{1}dx_{1}\wedge dx_{2} - \epsilon_{2}dx_{3}\wedge dx_{4}\).  Then we define the refined topological string to be the result of preforming the same Schwinger calculation in this modified background,
\begin{equation}
F_{ref\;top} = \sum_{\beta, s_{L}, s_{R}}\sum_{d=1}^{\infty}\sum_{j_{L}=-s_{L}}^{s_{L}}\sum_{j_{R}=-s_{R}}^{s_{R}}\frac{1}{d}\frac{(-1)^{2s_{L}+2s_{R}}N^{s_{L},s_{R}}_{\beta}(q/t)^{d\, j_{R}}(qt)^{d\, j_{L}}Q^{\beta d}}{(q^{d/2}-q^{-d/2})(t^{d/2}-t^{-d/2})}
\end{equation}
where \(q=e^{i \epsilon_{1}}\) and \(t=e^{-i \epsilon_{2}}\).  This gives a working definition of the refined topological string, and makes it clear that refinement captures much more information about the spin structure of the BPS spectrum.  In the rest of this paper, we will often refer to this picture as the ``first-quantized'' perspective, since the free-energy is related to counting of single BPS states, rather than the ``second-quantized'' perspective which we now introduce.

As explained in \cite{Hollowood:2003cv, Dijkgraaf:2006um}, the partition function, \(Z_{top} = \exp(F_{top})\) of the unrefined topological A-model on a Calabi-Yau threefold, \(X\), can be computed as a trace over the second-quantized Hilbert space of BPS states.  We take M-theory on \(\mathbb{C}^{2} \times S^{1} \times X\) and compute the five-dimensional trace,
\begin{equation}
Z_{\textrm{M-theory}}(X,q) = \textrm{Tr} \; (-1)^{F}q^{S_{1}-S_{2}}e^{-\beta H}
\end{equation}
where \(S_{1}\) and \(S_{2}\) are rotations in the \(z_{1}\) and \(z_{2}\) planes respectively, and \(\beta\) is the radius of the thermal circle.  This trace only receives contributions from BPS states (see Appendix \ref{sec:index} for details), and is equivalent to the geometry,
\begin{equation}
(X \times TN \times S^{1})_{q} \label{eqn:ordCS}
\end{equation}
where \(TN\) denotes the Taub-NUT space.  The Taub-NUT is twisted along the ``thermal'' \(S^{1}\) so that upon going around the \(S^{1}\) we have the rotation,
\begin{eqnarray}
z_{1} & \to & q z_{1} \\
z_{2} & \to & q^{-1}z_{2} \nonumber
\end{eqnarray}
where \(q=e^{ig_{s}}\) as before.  Note that the integrality structure of the Gopakumar-Vafa invariants in the graviphoton background computation is precisely what ensures that the result can be interpreted as a second-quantized trace.

This relation between M-Theory and the topological A-model extends to the \emph{open} topological string sector as follows \cite{Ooguri:1999bv, Aganagic:2009cg, Cecotti:2010fi, Cheng:2010yw}.  Consider the open topological A-model with \(N\) A-branes wrapping a special lagrangian 3-cycle, \(L\), inside a Calabi-Yau threefold, \(X\).  The corresponding M-theory partition function comes from wrapping \(N\) M5-branes on,
\begin{equation}
(L\times \mathbb{C} \times S^{1})_{q}
\end{equation}
where the \(\mathbb{C}\) is the cigar-shaped submanifold, \(\{z_{2}=0\}\), sitting inside the Taub-NUT space.  Note that the M5-brane partition function is given explicitly by the same index,
\begin{equation}
Z_{M5}(L,X,q) = \textrm{Tr} \; (-1)^{F}q^{S_{1}-S_{2}}e^{-\beta H}
\end{equation}
Note however, that now \(S_{2}\) has the interpretation of R-charge from the perspective of the \(M5\) brane while \(S_{1}\) corresponds to a rotation along the brane.  For more details on the indices that are relevant for the open and closed, refined and unrefined topological strings, see Appendix \ref{sec:index}.  

Now we would like to introduce a refined index for both the closed and open A-model.  Instead of  \ref{eqn:ordCS}, consider the refined M-Theory geometry,
\begin{equation}
(X \times TN \times S^{1})_{q,t}
\end{equation}
where upon going around the \(S^{1}\), the Taub-NUT space is twisted by,
\begin{eqnarray}
z_{1} & \to & q z_{1} \\
z_{2} & \to & t^{-1} z_{2} \nonumber
\end{eqnarray}
As explained in \cite{Aganagic:2011mi, CSRefined}, when \(t \neq q\) this configuration will break supersymmetry.  However, when \(X\) is a \emph{non-compact} Calabi-Yau threefold, M-Theory on \(X\) geometrically engineers a five-dimensional theory which has a conserved \(U(1)_{R}\subset SU(2)_{R}\) symmetry.  Then we can modify the construction to preserve supersymmetry by including an R-symmetry twist as we go around the \(S^{1}\).  When \(X\) is non-compact, this \(U(1)_{R}\) is actually realized geometrically by a Killing vector in the Calabi-Yau (see \cite{Nakayama:2011be} for a recent discussion of this symmetry).  This geometry then computes the index,
\begin{equation}
Z_{\textrm{refined top}}(X,q,t) \equiv Z_{\textrm{M-theory}}(X,q,t) = \textrm{Tr} \; (-1)^{F}q^{S_{1}-S_{R}}t^{S_{R}-S_{2}}e^{-\beta H} \label{eqn:refindex}
\end{equation}
and gives a definition of the refined closed topological string on \(X\).\footnote{It is important to emphasize that the refined topological string computes this protected spin character rather than simply counting the refined BPS multiplicities (as it would without the additional \(U(1)_{R}\) twist).  The refined BPS multiplicities themselves, \(N^{(j_{L},j_{R})}_{\beta}\), are not invariant under changes of the complex structure \cite{Dimofte:2009bv}, but the protected spin character \emph{is} invariant.  When the Calabi-Yau has no complex structure deformations, the naive and protected indices agree if no ``exotic'' BPS states (states with nonzero R-charge) are present in the spectrum.  The absence of such exotic BPS states in the four-dimensional field theory limit was conjectured in \cite{Gaiotto:2010be}.}

It is instructive to consider the two possible dimensional reductions along either the thermal \(S^{1}\) or the \(S^{1}\) of the Taub-NUT space, as studied in the unrefined case in \cite{Gaiotto:2005gf, Dijkgraaf:2006um}.  If we consider \(X\) to be noncompact, then this geometry engineers a five dimensional gauge theory and by reducing on the thermal \(S^{1}\) we precisely obtain the Nekrasov partition function of the gauge theory at \((i \epsilon_{1},i \epsilon_{2})=(\log q, -\log t)\) \cite{Nekrasov:2002qd}.

If we instead reduce along the Taub-NUT \(S^{1}\) we obtain IIA string theory on the geometry,
\begin{equation}
X \times \mathbb{R}^{3} \times S^{1}
\end{equation}
with a D6 brane wrapping \(X \times S^{1}\) and sitting at the origin of \(\mathbb{R}^{3}\).  Here it is helpful to recall some useful facts about the Taub-NUT geometry.  The geometry has a \(U(1)_{L} \times SU(2)_{R}\) isometry, which we have used above in the definition of the index.  Asymptotically, Taub-NUT looks like \(S^{1} \times \mathbb{R}^{3}\) and the \(U(1)_{L}\) isometry rotates the \(S^{1}\), while the \(SU(2)_{R}\) rotates the base geometry.  So upon dimensional reduction, the charge under \(U(1)_{L}\) becomes the \(D0\) charge, while the charge under \(U(1)_{R}\) becomes the spin in the base \(\mathbb{R}^{3}\).  Explicitly, \(\sqrt{qt}\) becomes the \(D0\) chemical potential, while \(\sqrt{q/t}\) is the chemical potential for a combination of spin and R-charge, and the index on the D6 brane can be rewritten as,
\begin{equation}
Z_{\textrm{refined top}}(X,q,t) = \textrm{Tr}_{D6} \; (-1)^{F}q_{1}^{Q_{0}}q_{2}^{2J_{3}-2S_{R}}
\end{equation}
where \(q_{1}=\sqrt{qt}\), \(q_{2}=\sqrt{q/t}\), \(Q_{0}\) is the \(D0\) brane charge, and \(J_{3}\) is the generator of the rotation group in \(\mathbb{R}^{3}\).  Upon setting \(q_{2}=1\), this gives the unrefined topological string, which is known to be equivalent to the topologically twisted theory living on a D6 brane wrapping X.  In the refined case, this computation of refined BPS states bound to one \(D6\) brane should be equivalent to the refined topological string.

It is natural to extend this construction to the open string case by inserting a stack of N M5-branes wrapping a special lagrangian \(L\) inside \(X\).  We will also require that \(L\) is fixed by the geometric \(U(1)_{R}\) killing vector in \(X\).  Then the full geometry wrapped by the \(M5\) branes is given by,
\begin{equation}
(L \times \mathbb{C} \times S^{1})_{q,t}
\end{equation}
where again \(\mathbb{C}\) denotes the locus \(\{z_{2}=0\}\) inside the Taub-NUT space.  Now we can compute the same index as in the closed case,
\begin{equation}
Z_{M5}(L,X,q,t) = \textrm{Tr} \; (-1)^{F}q^{S_{1}-S_{R}}t^{S_{R}-S_{2}}e^{-\beta H} \label{eqn:sunindex}
\end{equation}

This gives a definition of the refined open topological string theory on \(X\) in the presence of N refined A-branes wrapping \(L\).  Now recall that in the unrefined case, this M-theory partition function was related to ordinary Chern-Simons theory for the choice, \(X=T^{*}L\).  In the refined case, we do not yet have a path integral definition of the refined Chern-Simons theory, so it is natural to take this M-theoretic construction as the \emph{definition} of $SU(N)$ refined Chern-Simons theory,
\begin{equation}
Z_{ref\;CS}(L,q,t;\; SU(N)) := Z_{\textrm{N M5}}(L, T^{*}L, q, t)
\end{equation}

Again we can gain further insight by alternatively reducing the M-theory geometry on either the thermal \(S^{1}\) or the Taub-NUT \(S^{1}\).  Reducing on the thermal \(S^{1}\) we obtain the omega background in the presence of a surface operator.  Of course, it is worth noting that in some examples (\(X = T^{*}S^{3}\)), the ``geometrically engineered'' gauge theory in the bulk will be trivial and the only nontrivial dynamics will live on the surface operator.  However, more generally we will obtain a surface operator coupled to a gauge theory, with the Omega-deformed theory computing a coupled instanton-vortex partition function \cite{Dimofte:2010tz}.  The wall crossing behavior of such coupled 2d-4d systems has recently been studied in \cite{Gaiotto:2011tf}.

Alternatively, reducing on the Taub-NUT \(S^{1}\) we are left with a \(D4\) brane wrapping \(S^{1} \times \mathbb{R}_{+} \times L\) and ending on the D6 brane.  This is precisely the geometry considered by Witten in connection with Khovanov homology \cite{Witten:2011zz}.  The only difference is that here we compute an index by utilizing the additional \(U(1)_{R}\) symmetry, whereas Witten studies Khovanov homology directly, which \emph{cannot} be computed as an index as it only involves the gradings, \(S_{1}\) and \(S_{2}\).

\section{Orientifolds and M-Theory \label{sec:ofolds}}
Having reformulated refined topological string theory as a trace in M-theory, we now introduce orientifolds.  We will restrict to orientifolds that act as an anti-holomorphic involution, \(I: X \to X\), on the Calabi-Yau, including both possibilities of \(I\) having fixed points or acting freely.  

In physical string theory, orientifolds are defined by specifying a simultaneous involution on the spacetime, \(I\), and orientation reversal, \(\Omega\), on the worldsheet.  This definition can be extended to the ordinary topological string \cite{Acharya:2002ag}, by starting with the covering space, \(\Sigma\), of an unorientable worldsheet.  Then the A-model will localize on holomorphic maps, \(\phi : \Sigma \to X\) satsifying the additional constraint that  \(I \circ \phi = \phi \circ \Omega\).  Mathematically, this means that in the presence of an orientifold, the topological string is counting holomorphic maps which are \(\mathbb{Z}_{2}\)-equivariant.  Note that since orientation reversal on the worldsheet is anti-holomorphic, this constraint only makes sense if \(I\) acts anti-holomorphically on \(X\) as we have required.  To fully specify an orientifold both in the physical and the topological string, we must make a choice of the sign of the cross-cap amplitude.  In this paper we will restrict to the negative sign case (for open strings, this leads to an \(SO(N)\) gauge group), as this choice simplifies the lift to M-theory.

This gives a working worldsheet definition of unoriented topological strings, but it was further conjectured in \cite{SinhaVafa, Bouchard:2004iu, Bouchard:2004ri, Walcher:2007qp} that the unoriented topological string can be rewritten by counting single-particle BPS states.  There it was argued that the orientifolded topological string computes terms in the low energy effective action of IIA string theory on \(X \times \mathbb{C}^{2}\), with the involution acting simultaneously as \(I\) on \(X\) and as a reflection on two of the spacetime coordinates, \((z_{1}, z_{2}) \to (z_{1}, -z_{2})\).  For future reference, we refer to this combined action as \(\tilde{I}\).  In the case when \(I\) has a fixed locus, \(L\), this corresponds to wrapping an \(O4\) plane on \(L \times \mathbb{C}\).  As in the ordinary case, these terms in the effective action should also be computable by introducing a self-dual graviphoton background and studying the contribution of wrapped branes\footnote{ In order to show this more rigorously, we would need to determine exactly which terms in the orientifolded \(\mathcal{N}=1\) low energy effective action are computed by the topological string.  Then by integrating out wrapped brane contributions to these terms, we could derive this integrality structure.  This was done at genus zero in \cite{Acharya:2002ag}, but it would be interesting to study the higher genus amplitudes in more detail.}.  This results in the free energy of the closed topological A-model taking the general form,

\begin{eqnarray}
\mathcal{F}(X/I, g) & = & \frac{1}{2}\mathcal{F}(X,g) + \mathcal{F}(X/I,g)_{unor} \\
& = & \frac{1}{2}\sum_{d=1}^{\infty}\sum_{g=0}\sum_{\beta}\frac{1}{d}\frac{N^{g}_{\beta}}{(q^{d/2}-q^{-d/2})^{2-2g}}Q^{\beta d}
\label{eqn:forient} \\
 & & + \sum_{d\; odd/even}\sum_{g=0}\sum_{\beta}\frac{1}{d}\frac{N^{g,c=1}_{\beta}}{(q^{d/2}-q^{-d/2})^{1-2g}}Q^{\beta d} \nonumber \\
 & & + \sum_{d\; odd/even}\sum_{g=0}\sum_{\beta}\frac{1}{d} N^{g,c=2}_{\beta}(q^{d/2}-q^{-d/2})^{2g}Q^{\beta d} \nonumber
\end{eqnarray}
where the \(N^{g,c}_{\beta}\) are integers that count branes wrapping the cycle, \(\beta \in H_{2}(X/I,\mathbb{Z})\).\footnote{In \cite{Walcher:2007qp, Krefl:2009md, Krefl:2009mw}, it was found experimentally that for some orientifold actions with a non-toric fixed locus, the integrality of the BPS states only holds when a D-brane is introduced wrapping the same locus as the orientifold fixed plane.  For example, such a D-brane must be introduced in the local \(\mathbb{P}^{2}\) geometry when the orientifold fixed plane wraps a real line bundle over the real locus, \(\mathbb{RP}^{2}\).  In such cases, it was argued that only the combined contribution from open strings and unoriented strings gives the expected BPS integrality structure, and the resulting integers are known as \emph{real} Gopakumar-Vafa invariants.  Although we will not discuss these cases explicitly, we will find that our M-theory integrality conjecture, explained in the following section, applies to these geometries as well.  In fact, the equivalent of our strong integrality property was originally stated for real GV invariants by Walcher in \cite{Walcher:2007qp}.}  In the unoriented sector, the sums are over either odd \(d\) or even \(d\), depending on the details of the orientifold action \(I\).  Note that the orientifold action explicitly breaks the spacetime rotational symmetry, \(SU(2)_{L} \times SU(2)_{R}\) down to \(U(1)_{1} \times U(1)_{2}\).  Because of this breaking, it is no longer guaranteed that the BPS states will come in full spin multiplets.  However, the \((q^{d/2}-q^{-d/2})^{2g}\) factors come from the moduli space of flat connections on a wrapped \(D2\) brane, even if the brane is wrapping an unorientable cycle.  As explained below, this moduli space generically takes the form of \(T^{2g}\) so we expect that the BPS contributions will continue to sit in full \(SU(2)_{L}\) multiplets. 

To understand this conjecture physically, note that except for the factor of \(\frac{1}{2}\), the first term is the same as the free energy for the topological string on \(X\) in the absence of an orientifold.  This term counts BPS states that wrap oriented cycles in the \(X\) and are free to move in the four spacetime directions.  Since these states can move in \(\mathbb{C}^{2}\), we obtain a factor of \((q^{d/2}-q^{-d/2})^{2}\) in the denominator.  Although we do not know of a convincing target space interpretation for the factor of \(\frac{1}{2}\), \cite{SinhaVafa} argued that it arises on the worldsheet from dividing by the orientation reversal symmetry, \(\Omega\).

Now we turn to the unoriented contributions to the free energy.  These come from branes wrapping BPS cycles in \(X/I\).  These cycles will lift to surfaces with boundaries in the covering space \(X\) (if they did lift to closed BPS two-cycles, then this would be an oriented rather than an unoriented contribution).  Now recall that the orientifold acts on the full geometry as \(I\) on \(X\) and as \(z_{2} \to - z_{2}\) in two of the spacetime dimensions.  When we take into account this full orientifold action, in order to get a genuine closed BPS state with no boundaries, it is necessary that the brane sits at \(z_{2} = 0\).  This means that the unoriented branes effectively propagate in only two noncompact dimensions, so the denominator of the unoriented contributions will give only one power of \((q^{d/2}-q^{-d/2})\).

The first unoriented contribution to the free energy comes from D-branes wrapping unorientable surfaces with one crosscap.  Recall that a crosscap is inserted into a surface by cutting open a hole and identifying antipodal points on the boundary.  The lowest order contribution comes from curves with genus \(0\) and one crosscap, which have the topology of \(\mathbb{RP}^{2}\).  

To understand the structure in more detail, it is helpful to recall that the first homology of a genus g surface with one crosscap, \(\Sigma^{1}_{g}\), is given by,
\begin{equation}
H_{1}(\Sigma^{1}_{g}, \mathbb{Z}) = \mathbb{Z}^{2g}\oplus \mathbb{Z}/2\mathbb{Z}
\end{equation}
 This tells us that the continuous part of the moduli space of flat connections on \(\Sigma^{1}_{g}\) is generically given by \(T^{2g}\) as in the original analysis of Gopakumar and Vafa.  This accounts for the \((q^{d/2}-q^{-d/2})^{2g}\) factor in the numerator of the \(n^{g, c=1}_{Q}\) term.  Because of the \(\mathbb{Z}/2\mathbb{Z}\) factor, we also have the option of turning on one unit of discrete flux, which corresponds to dissolving half a unit of \(D0\) brane charge in the wrapped \(D2\) brane.  As argued in \cite{SinhaVafa}, this half unit of \(D0\) brane flux will shift \(t \to t + \pi i\), or equivalently, \(Q^{d} \to (-1)^{d}Q^{d}\).  Depending on how \(I\) acts, this may change the fermion number assignment and introduce an additional overall minus sign.  Adding together both choices of discrete flux, we obtain the cancellation that accounts for summation over only even/odd \(d\).

Finally, we can do the same analysis on the terms with two crosscaps.  It is helpful to recall that the first homology of a genus \(g\), \(c=2\) surface is given by,
\begin{equation}
H_{1}(\Sigma^{2}_{g}, \mathbb{Z}) = \mathbb{Z}^{2g+1}\oplus \mathbb{Z}/2\mathbb{Z}
\end{equation}
As in the single crosscap case, this means that a genus \(g\) surface will have a moduli space of flat connections given by, \(T^{2g+1}\), which leads to a factor of \((q^{d/2}-q^{-d/2})^{2g+1}\).  Note that for the genus \(0\) surface with two crosscaps, the Klein bottle, this factor in the numerator cancels the factor in the denominator from moving in two noncompact dimensions.  Again, the additional torsion factor \(\mathbb{Z}/2\mathbb{Z}\) accounts for the restriction to even/odd \(d\).  Note that we have exhausted all possible unorientable surfaces, since in the presence of an additional crosscap, two crosscaps can be traded for a handle.  Although we will not discuss it in detail here, by similar arguments we expect the open topological string in the presence of orientifolds to possess nice integrality properties from counting BPS states ending on \(D4\) branes.

Now that we have given the first-quantized BPS interpretation, it is natural to conjecture that, as in the oriented case, the partition function of the topological string, \(Z_{top}\) is computed by a second quantized trace in M-theory.  As the first step, we must explain how the IIA orientifold action lifts to M-theory.  

Of course, since there is no worldsheet in M-theory, the \(IIA\) orientifolds must lift to orbifolds in eleven dimensions.  The \(IIA\) orientifold, with \(\tilde{I}\) acting on the spacetime and \(\Omega\) acting on the worldsheet, lifts to the M-theory orbifold by \(\tilde{I}\).  To fully specify the orbifold, we need to explain how the three-form, \(C_{(3)}\) is affected by the orbifold action.  As explained in \cite{Witten:1995em, Dabholkar:1996pc, deBoer:2001px}, \(\tilde{I}\) acts as \(C_{(3)} \to -C_{(3)}\).  One simple way to see this is to recall that the low-energy action of M-theory contains a Chern-Simons-like term,
\begin{equation}
\int C\wedge dC \wedge dC
\end{equation}
All terms in the action should be invariant under the orbifold action, and since \(\tilde{I}\) is orientation-reversing, the Chern-Simons term will only be invariant if \(C \to -C\).

In the case where \(I\) has a fixed locus, the string theoretic \(O4^{-}\) plane will lift to the six-dimensional fixed locus of the M-theory orbifold, which we refer to as an \(MO5\)-plane.  However, when \(I\) acts freely, the orientifold action lifts to M-theory on the unorientable, smooth geometry, \(\big(X \times \mathbb{C}\big)/\tilde{I} \times \mathbb{C} \times S^{1}\).  It is important to remember that the action of \(\tilde{I}\) multiplies \(C\) by \(-1\), so that in this unorientable M-theory geometry, \(C\) is actually a \emph{twisted} three-form rather than an ordinary three-form \cite{Witten:1998xy}.

Having explained the relevant objects in M-theory, we are now ready to study the topological string.  We would like to compute the M-theory trace by taking the geometry,
\begin{equation}
\Big(\mathbb{C}_{1} \times (\mathbb{C}_{2} \times X)/\tilde{I} \times S^{1}\Big)_{q}
\end{equation}
where as before, \((\cdots)_{q}\) indicates that as we go around the \(S^{1}\), we rotate the \(\mathbb{C}_{1} \times \mathbb{C}_{2}\) by \((z_{1},z_{2}) \to (q z_{1}, q^{-1}z_{2})\).  Note that the orbifold takes \(\theta_{2} \to \theta_{2} + \pi\), while the the \(S^{1}\) twist takes \(\theta_{2} \to \theta_{2} - g_{s}\), so these operations commute with each other, as they must in order to define a sensible geometry.  This defines the closed topological string in the presence of orientifolds by representing it as an M-theory trace.  Similarly, the open topological string is computed exactly as before by introducing \(M5\) branes wrapping special lagrangain submanifolds in \(X\), but now in the presence of orientifolds.

Now we are finally ready to give a definition of the refined topological string in the presence of orientifolds.  Although it may be possible to give a definition from the first-quantized graviphoton picture, we have found that refinement is simpler and clearer in the second-quantized picture.

As in the oriented case, to refine we simply need to compute a more general M-theory partition function.  This leads us to consider the geometry,
\begin{equation}
\Big(\mathbb{C}_{1} \times (\mathbb{C}_{2} \times X)/\tilde{I}  \times S^{1}\Big)_{q,t}
\end{equation}
where as in the oriented case, the \(\mathbb{C}^{2}\) geometry is rotated as we go around the \(S^{1}\) ,
\begin{eqnarray}
z_{1} & \to & q z_{1} \\
z_{2} & \to & t^{-1}z_{2} \nonumber
\end{eqnarray}
where we also must include a rotation by the \(U(1)_{R}\) to preserve supersymmetry \footnote{For this to make sense, we must also require that the orientifold action \(I\) is compatible with the isometry generating the \(U(1)_{R}\) symmetry.}.  Then we define the refined topological string in the presence of orientifolds to be equal to this M-theory partition function,
\begin{equation}
Z_{ref\;closed}(X/I;q,t) = \textrm{Tr}_{\textrm{M-Theory}}\; (-1)^{F}q^{S_{1}-S_{R}}t^{S_{R}-S_{2}}e^{-\beta H}
\end{equation}
Similarly, the refined open topological string is defined by introducing \(M5\) branes wrapping special lagrangians and computing the same trace.

Before moving on to explicit computations, it is interesting to look at the general form of this M-theory partition function.  We will focus on the closed case, although the open case is completely analogous.  There are two types of BPS states contributing to the partition function.  The first contribution comes from BPS M2-branes wrapping closed, orientable two-cycles.  Each BPS state gives a field in four dimensions, \(\Phi(z_{1},\bar{z}_{1},z_{2}, \bar{z}_{2})\), and this field has additional excitations which are BPS, provided that \(\Phi\) depends holomorphically on \(z_{1}\) and \(z_{2}\).  It is natural to decompose \(\Phi\) into modes as,
\begin{equation}
\Phi = \sum_{l_{1}, l_{2}}\alpha_{l_{1}, l_{2}}z_{1}^{l_{1}}z_{2}^{l_{2}}
\end{equation}

In the full M-theory partition function, we must include the contributions of these modes, which carry angular momentum in the \(\mathbb{C}_{1}\) and \(\mathbb{C}_{2}\) planes.  However, here it is important to remember that the M-theory orbifold acts nontrivially on the spacetime as \(z_{2} \to -z_{2}\).  The field, \(\Phi\), should have a well-defined transformation property under this orbifold, so that \(\Phi(z_{1},-z_{2}) = \pm \Phi(z_{1},z_{2})\), where the choice of \(\pm\) is related to how the orbifold acts in \(X\).  This means that we should only keep either the even or odd modes of \(l_{2}\). 

Thus, the contribution to \(Z_{ref \; top}\) from an \(M2\) brane with intrinsic spin, \((m_{1},m_{2})\), wrapping a two-cycle in the class \(\beta \in H_{2}(X,\mathbb{Z})\) takes the form,
\begin{equation}
\prod_{l_{1}, l_{2}=1}^{\infty}\Big(1-q^{m_{1}+l_{1}}t^{m_{2}+2l_{2}}Q^{\beta}\Big)
\end{equation}

The second type of contribution comes from M2 branes ending on the orbifold (when \(I\) has fixed points) or M2 branes wrapping unorientable cylces (when \(I\) does not have fixed points).  These \(M2\) branes are frozen at the fixed locus, \(z_{2}=0\) (since otherwise they would have a boundary).  This means that they make a contribution in the form of a quantum dilogarithm,
\begin{equation}
\prod_{l_{1} =1}^{\infty}\Big(1-q^{m_{1}+l_{1}}t^{m_{2}}Q\Big)
\end{equation}
Putting these contributions together, we find that \(Z\) takes the form,
\begin{equation}
Z_{ref}(X/I; q,t) = \prod_{l_{1}, l_{2}=1}^{\infty}\Big(1-q^{m_{1}+l_{1}}t^{m_{2}+2l_{2}}Q^{\beta}\Big)^{M^{(m_{1},m_{2})}_{\beta}} \prod_{l_{1} =1}^{\infty}\Big(1-q^{m_{1}+l_{1}}t^{m_{2}}Q\Big)^{\widetilde{M}^{(m_{1}, m_{2})}_{\beta}} \label{eqn:zref}
\end{equation}
where the partition function is completely determined by the integer invariants, \(M^{(m_{1},m_{2})}_{\beta}\) which counts oriented \(M2\) branes and \(\widetilde{M}^{(m_{1}, m_{2})}_{\beta}\), which counts the unoriented branes in \(X/I\).

\section{A New Integrality Conjecture from M-Theory \label{sec:integrality}}
It is interesting to take equation \ref{eqn:zref} and return to the unrefined case, \(t=q\).  We then have integrality properties following from both the second-quantized M-theory picture and from the first-quantized Gopakumar-Vafa picture.  When no orientifolds are present, these integrality properties are equivalent.  More explicitly, the first-quantized integer invariants appearing in the free energy, \(F\), guarantee the integrality in the second-quantized partition function, \(Z=e^{F}\).

However, once we introduce orientifolds, the first-quantized structure (equation \ref{eqn:forient}) includes overall factors of \(1/2\) in the free energy.  This poses a serious problem since such half-integers will result in terms of the form \(\sqrt{1-q^{n}Q^{\beta}}\) in the partition function.  Such terms have no sensible interpretation as counting second-quantized states in a free Fock space.

Thus, if the topological string partition function can truly be defined as an index in M-theory, we require a stronger integrality constraint on the Gopakumar-Vafa invariants \(N^{g,c}_{\beta}\) appearing in the free energy.  To see precisely which constraints are needed, it is helpful to rewrite the unoriented contributions to the free energy as \footnote{If we instead looked at the case of an even sum over \(d\), only the second term would appear, with opposite sign.  Since the second term is the crucial one in our analysis, precisely the same conclusions hold regardless of whether the sum is over even or odd \(d\).}
\begin{eqnarray}
& & \sum_{d\;\; odd}\sum_{g=0}^{\infty}\sum_{\beta}\frac{1}{d}\frac{N^{g,c}_{\beta}}{(q^{d/2}-q^{-d/2})^{1-2g}}Q^{\beta d} \\
& = &  \sum_{d=1}^{\infty}\sum_{g=0}^{\infty}\sum_{\beta}\frac{1}{d}\frac{N^{g,c}_{\beta}}{(q^{d/2}-q^{-d/2})^{1-2g}}Q^{\beta d} - \sum_{d=1}^{\infty}\sum_{g=0}^{\infty}\sum_{\beta}\frac{1}{2d}\frac{N^{g,c}_{\beta}}{(q^{d}-q^{-d})^{1-2g}}Q^{2\beta d} \nonumber
\end{eqnarray}
This form makes it clear that half-integers can appear in both the oriented and unoriented sector.  Of course, the simplest way to guarantee integrality is if all the Gopakumar-Vafa invariants, \(N^{g}_{\beta}\) and \(N^{g, c}_{\beta}\), are even.  However, generically, we do not expect this to be the case.  Instead, integrality can be achieved more generally if the half-integer contributions in the oriented and unoriented sector partially cancel each other to give integers.

To see explicitly how this occurs, let us focus on the oriented genus \(g\) contribution of an M2 brane wrapping \(\beta\), and take \(g\) to be even.  Now we can combine this with the one-crosscap (\(c=1\)), genus g amplitude for curves wrapping \(\beta/2\).  Grouping the half-integer pieces of each term gives,
\begin{equation}
\frac{1}{2d}\Bigg(\frac{N^{g}_{\beta}(q^{d/2}-q^{-d/2})^{2g}}{(q^{d/2}-q^{-d/2})^{2}}-\frac{N^{g/2, c=1}_{\beta/2}(q^{d}-q^{-d})^{g}}{(q^{d}-q^{-d})}\Bigg)Q^{\beta d}
\end{equation}
This can be rewritten as,
\begin{eqnarray}
\frac{Q^{\beta d}(q^{d/2}-q^{-d/2})^{g}}{2d(q^{d/2}-q^{-d/2})(q^{d}-q^{-d})} & & \Bigg(N^{g}_{\beta}\Big(q^{d/2}-q^{-d/2}\Big)^{g}(q^{d/2}+q^{-d/2}) \\
& &-N^{g/2, c=1}_{\beta/2}\Big(q^{d/2}+q^{-d/2}\Big)^{g}(q^{d/2}-q^{-d/2})\Bigg) \nonumber
\end{eqnarray}
Now observe that the q-factors in front of \(N^{g}_{\beta}\) and \(N^{g/2, c=1}_{\beta/2}\) are identical except for signs.  When we expand these terms out, we will obtain a polynomial in \(q\) whose coefficients are multiples of either \(N^{g}_{\beta}+ N^{g/2, c=1}_{\beta/2}\) or \(N^{g}_{\beta} - N^{g/2, c=1}_{\beta/2}\).  So to cancel the factor of \(1/2\) in front of the expression, we require that the Gopakumar-Vafa invariants are both even or both odd.

We can repeat the same analysis by combining the odd genus contributions with the \(c=2\) terms.  Putting everything together, we find that the second quantized M-theory interpretation of the topological string implies the strong integrality property,
\begin{equation}
N^{g}_{\beta} \equiv \left\{ \begin{array}{ll}
N^{g/2, c=1}_{\beta/2} & \textrm{for \(g\) even} \\
N^{(g-1)/2, c=2}_{\beta/2} & \textrm{for \(g\) odd}
\end{array} \right. \textrm{\;\; (mod 2)}
\end{equation}
Note that in come cases, there cannot be any BPS state wrapping the class \(\beta/2\).  In this case, the strong integrality property tells us that \(N^{g}_{\beta}\) itself must be even.  We have verified that this integrality property is satisfied for all of the geometries studied in \cite{Bouchard:2004iu, Walcher:2007qp, Krefl:2009md}, including orientifolds of the conifold, the quintic, local \(\mathbb{P}^{2}\), and more general toric geometries.  

An equivalent observation was made by Walcher for the case of certain real orientifolds, with a D-brane wrapping the fixed locus \cite{Walcher:2007qp}.  Walcher argued that if we pretend the moduli space of oriented wrapped branes is a collection of points, then \(I\) acts on this set.  The points that are fixed by \(I\) are precisely the BPS states wrapping the fixed real locus, while the other points must come in pairs that are exchanged by \(I\).  This implies that the ``real'' (unoriented) and oriented Gopakumar-Vafa invariants must differ by a multiple of two.  Here we have generalized this statement to include arbitrary orientifolds, and have given it a natural M-theory interpretation.  

At this point we could go one step further and obtain expressions for the natural M-theoretic integer invariants, \(M^{s}_{\beta}\) and \(\widetilde{M}^{s}_{\beta}\), in terms of the \(N^{g,c}_{\beta}\) by expanding out the polynomials in q.  It is interesting to note that these variables are complimentary, in the sense that the \(M, \widetilde{M}\) variables make integrality manifest, but obscure the organization of invariants in full spin multiplets.  In contrast, the natural free energy variables, \(N^{g,c}_{\beta}\) make manifest the spin multiplet structure, while hiding the full integrality properties.

We can also consider the integrality structure when we have branes wrapping a special Lagrangian 3-cycle, \(L\), inside \(X\).  When the orientifold action, \(I\), has fixed points, we will assume that the new branes do not sit at the fixed locus.  As explained in \cite{Bouchard:2004ri}, the open topological string will receive contributions from an oriented and an unoriented sector and the free energy will include half-integer contributions.  As discussed above, the partition function, \(Z=e^{F}\), of the open topological string can be written as a second-quantized trace in M-theory counting \(M2\) branes ending on \(M5\) branes wrapping \(L\).  For this M-theory interpretation to hold, the half-integers in the free energy must cancel, leading us to a new integrality condition in the open case. 

To explain this in more detail, we begin with the first-quantized form of the open string free energy conjectured in \cite{Bouchard:2004ri},

\begin{equation}
F(X/I, V) = \frac{1}{2}\sum_{R_{1}, R_{2}}\sum_{d=1}^{\infty}\frac{1}{d}f^{cov}_{R_{1},R_{2}}(q^{d},Q^{d})\textrm{Tr}_{R_{1}\otimes R_{2}}V^{d} - \sum_{R}\sum_{d\;odd/even}\frac{1}{d}f^{unor}_{R}(q^{d},Q^{d})\textrm{Tr}_{R}V^{d}
\end{equation}

The first term comes from oriented open strings and can be computed by going to the covering space.  Since the branes are not fixed by \(I\), in the covering space there will be branes wrapping both \(L\) and its image under the involution, \(I(L)\).  However, since these two stacks of branes are related by \(I\), their open string moduli must be equal to each other, giving the trace \(\textrm{Tr}_{R_{1}}V\cdot\textrm{Tr}_{R_{2}}V = \textrm{Tr}_{R_{1}\otimes R_{2}}V\).  The second term comes from unoriented strings, with the sum over \(d\) restricted to odd or even positive integers depending on the orientifold action, as in the closed case.

It is convenient to rewrite the oriented piece as,
\begin{equation}
F^{or}(X/I, V) = \frac{1}{2}\sum_{R}\sum_{d=1}^{\infty}\frac{1}{d}\Bigg(\sum_{R_{1}, R_{2}}f^{cov}_{R_{1},R_{2}}(q^{d},Q^{d})N^{R}_{R_{1}, R_{2}}\Bigg)\textrm{Tr}_{R}V^{d}
\end{equation}
where \(N^{R}_{R_{1},R_{2}}\) is the Littlewood-Richardson coefficient for decomposing the tensor product \(R_{1} \otimes R_{2}\).  We can also rewrite the unoriented piece (assuming the sum is over odd \(d\)) as,
\begin{equation}
F^{unor}(X/I, V) = - \sum_{R}\sum_{d=1}^{\infty}\frac{1}{d}f^{unor}_{R}(q^{d},Q^{d})\textrm{Tr}_{R}V^{d} + \sum_{R}\sum_{d=1}^{\infty}\frac{1}{2d}f^{unor}_{R}(q^{2d},Q^{2d})\sum_{R'}c_{2;R}^{R'}\textrm{Tr}_{R'}V^{d}
\end{equation}
where \(c_{2\;R}^{R'}\) is the coefficient of the second Adams Operation defined by, 
\begin{equation}
\textrm{Tr}_{R}(V^{2}) = \sum_{R'} c_{2;R}^{R'}\textrm{Tr}_{R'}(V)
\end{equation}

It is natural to combine the half-integer pieces in the oriented and unoriented amplitudes to give,
\begin{equation}
\frac{1}{2}\sum_{R}\sum_{d=1}^{\infty}\frac{1}{d}\Bigg(\sum_{R_{1}, R_{2}}N^{R}_{R_{1}, R_{2}}f^{or}_{R_{1}, R_{2}}(q^{d}, Q^{d}) + \sum_{R'}c_{2;R'}^{R} f^{unor}_{R'}(q^{2d},Q^{2d})\Bigg)\textrm{Tr}_{R}V^{d} \label{eqn:openunor}
\end{equation}
We can expand the \(f\) functions as,
\begin{eqnarray}
f^{cov}_{R_{1}, R_{2}}(q,Q) & = & \sum_{\beta, s}N_{(R_{1}, R_{2}), \beta, s}Q^{\beta}q^{s} \\
f^{unor}_{R}(q,Q) & = & \sum_{\beta, s}\widetilde{N}_{R,\beta,s}Q^{\beta}q^{s}
\end{eqnarray}
where \(\widetilde{N}_{R,\beta,s}\) and \(N_{(R_{1}, R_{2}), \beta, s}\) are integers counting BPS states with spin \(s\), wrapping the relative homology class, \(\beta \in H_{2}(X,L)\), and in representation \(R\).  Then the absence of half-integers in the free energy imposes the condition,
\begin{equation}
\sum_{R_{1}, R_{2}}N^{R}_{R_{1},R_{2}} N_{(R_{1}, R_{2}),\beta,s} \equiv \sum_{R'}c^{R}_{2;R'}\widetilde{N}_{R',\beta/2,s/2}\qquad \textrm{(mod 2)}
\end{equation}
It is important to note that the integers \(\widetilde{N}_{R,\beta, s}\) and  \(N_{(R_{1}, R_{2}),\beta, s}\) that we have used above are not the most fundamental BPS invariants.  To exhibit the full BPS structure it is necessary to include the structure of the moduli space of flat connections and geometric deformations of an open \(D2\) brane \cite{Labastida:2000yw}.  In general this structure is more complicated and involves the Clebsch-Gordon coefficients of the symmetric group.  For simplicity, we focus on the case of \(R=\tableau{1}\), which leads to a simple integrality constraint on the fundamental BPS invariants, \(\widehat{N}_{R,\beta,g}\).

We start by expanding out Equation \ref{eqn:openunor} in traces over \(V\),
\begin{eqnarray}
 = \frac{1}{2}(f^{cov}_{\tableau{1},\cdot}+f^{cov}_{\cdot,\tableau{1}})\textrm{Tr}_{\tableau{1}}V & + & \Bigg(\frac{1}{2}\Big(f^{cov}_{\tableau{2},\cdot}(q,Q)+f^{cov}_{\cdot,\tableau{2}}(q,Q)\Big)+\frac{1}{2}\Big(f^{cov}_{\tableau{1},\;\cdot}(q^{2},Q^{2})+f^{cov}_{\;\cdot,\;\tableau{1}}(q^{2},Q^{2})\Big) \nonumber \\
 & & + f^{cov}_{\tableau{1},\tableau{1}}(q,Q) + f^{unor}_{\tableau{1}}(q^{2},Q^{2}) \Bigg)\textrm{Tr}_{\tableau{2}}V + \cdots
\end{eqnarray}
Now since \(I\) exchanges the two stacks of branes in the covering space, it follows that \(f^{cov}_{R,\cdot} = f^{cov}_{\cdot, R}\).  Therefore, the \(Tr_{\tableau{1}}V\) term and the first two terms of \(Tr_{\tableau{2}}V\) do not contribute any half integers, but the last two terms could.  Imposing integrality leads to the condition,
\begin{equation}
f_{\tableau{1}, \tableau{1}}^{or}(q,Q) \equiv f_{\tableau{1}}^{unor}(q^{2},Q^{2}) \qquad (\textrm{mod }2)
\end{equation}
Since we are working with the relatively simple \(\tableau{1}\) representation, these \(f\) functions can be expanded in terms of the fundamental BPS invariants,
\begin{eqnarray}
f^{or}_{\tableau{1},\tableau{1}}(q,Q) & = & \sum_{\beta, g}\widehat{N}_{(\tableau{1},\tableau{1}), g, \beta}(q^{1/2}-q^{-1/2})^{2g}Q^{\beta} \\
f^{unor}_{\tableau{1}}(q,Q) & = & \sum_{\beta, g}\widehat{N}^{c=1}_{\tableau{1},g,\beta}(q^{1/2}-q^{-1/2})^{2g}Q^{\beta}+\sum_{\beta, g}\widehat{N}^{c=2}_{\tableau{1}, g, \beta}(q^{1/2}-q^{-1/2})^{2g+1}Q^{\beta}
\end{eqnarray}
The invariant \(\widehat{N}^{c}_{R,g,\beta}\) counts BPS states of genus \(g\) with \(c\) crosscaps wrapping the class \(\beta\) and in representation \(R\).  We can group terms exactly as we did in the closed case, and we find the integrality condition,
\begin{equation}
\widehat{N}_{(\tableau{1},\tableau{1}), g, \beta} \equiv \left\{ \begin{array}{ll}
\widehat{N}^{c=1}_{\tableau{1}, g/2, \beta/2} & \textrm{for \(g\) even} \\
\widehat{N}^{c=2}_{\tableau{1}, (g-1)/2, \beta/2} & \textrm{for \(g\) odd}
\end{array} \right. \textrm{\;\; (mod 2)}
\end{equation}

It was argued in \cite{Marino:2009mw}, that when \(X\) is the resolved conifold and \(L=\mathcal{L}_{\mathcal{K}}\) is the special lagrangian corresponding to a knot \(\mathcal{K}\), then the composite BPS invariants, \(\widehat{N}_{(R_{1}, R_{2}),g,\beta}(\mathcal{K})\) are related to the HOMFLY polynomial in the composite representation \((R_{1}, R_{2})\).  By making use of this connection, the BPS invariants have been computed for many knots in \cite{Marino:2009mw, Paul:2010qu, Paul:2010wr}.  Using this data, we have explicitly verified our strong integrality conjecture for the unknot, trefoil, and \(T(2,5)\) knots.  It would be interesting to perform these checks for more complicated geometries.

In this section we have focused on unrefined amplitudes.  In the case of refined amplitudes, we do not yet have a first-quantized definition, but we still demand integrality from the second quantized M-theory index that defines the theory.  In Section \ref{sec:largen}, we will test this integrality by studying the large N limit of \(SO(2N)\) refined Chern-Simons theory, and interpreting the result as refined closed strings on the resolved conifold.

\section{Open Strings from Refined Chern-Simons Theory \label{sec:refinedcs}}

Now that we have given a definition of the refined topological string for oriented geometries and in the presence of orientifolds, we would like to solve the theory.  In the case of open, refined topological strings, this was done in \cite{CSRefined} by carefully analyzing the contributions of BPS states to the refined index.  As explained above, if we choose our Calabi-Yau to be the cotangent bundle over a three-manifold, \(T^{*}M\), with branes wrapping \(M\), then we expect that the open refined string theory should reduce to a three-dimensional field theory living on \(M\).  This theory is a refined version of Chern-Simons theory, and was defined for oriented strings in \cite{CSRefined}.  The crucial idea \cite{CSRefined} used to compute the M-theory indices was to cut up $M$ (along with \(T^{*}M\)) into pieces on which one can solve the theory explicitly, and glue the pieces back together. In this way, the $S$ and $T$ matrices of refined $SU(N)$ Chern-Simons theory could be deduced from M-theory.

In the unoriented case, the natural anti-holomorphic involution of \(T^{*}M\) is given by reversing the fibers of the cotangent bundle, so that \(p_{i} \to -p_{i}\).  This is equivalent to inserting an orientifold plane that wraps \(M\).  As is standard in both physical and topological string theory \cite{SinhaVafa}, the inclusion of an orientifold plane (with the appropriate choice of crosscap sign) simply changes the gauge group from \(SU(N)\) to \(SO(2N)\). We can now follow the rest of the steps from \cite{CSRefined} in the present context. Taking $M$ to be a solid torus, the refined M-theory index can be computed explicitly in the presence of orientifold action. 
Instead of going through this here, we refer the reader to Appendix \ref{sec:twist} for a detailed discussion of the derivation in the \(SO(2N)\) case. We will simply state the answer. We find that the \(SO(2N)\) theory is solved, analogously to the \(SU(N)\) case, by replacing characters of  \(SO(2N)\) with D-type Macdonald polynomials associated to the root system of \(SO(2N)\). 

Moreover, in Appendix \ref{sec:twist}, we also extend this to refined Chern-Simons theory with any simply-laced gauge group.  Although there is no known brane realization of the \(E_{a}\) gauge groups, the same analysis still works in the corresponding six-dimensional $(2,0)$ theory. In the appendix, we also give an explicit example of the connection with three dimensional field theory when \(M=\mathbb{R}^{2} \times S^{1}\).
 
In this section we describe the refined Chern-Simons theory in detail from the perspective of topological field theory.  This perspective will be especially helpful when we begin studying knot invariants in Section \ref{sec:knots}.  Although a Lagrangian for refined Chern-Simons theory is not yet known, we can still define the theory by describing its amplitudes on simple geometries.

\subsection{Refined Chern-Simons as a Topological Field Theory}

Recall that any three-dimensional topological field theory should assign a number to any closed three manifold, \(Z(M)\), and an element of the appropriate Hilbert space to any three manifold with boundary, \(\Psi(M) \in \mathcal{H}_{\Sigma}\), where \(\Sigma=\partial M\).  Further, diffeomorphisms that act geometrically on the boundary should be represented by unitary operators acting on the Hilbert space, \(\mathcal{O}: \mathcal{H}_{\Sigma} \to \mathcal{H}_{\Sigma} \).

The refined Chern-Simons theory can be thought of as a restricted topological field theory in three-dimensions.  It is only well-defined on Seifert three-manifolds, M, to which it assigns a number, \(Z(M)\).  This restriction can be understood by recalling the definition of open refined string theory, which requires the existence of a \(U(1)\) isometry on \(M\).  The refined Chern-Simons theory also assigns an element of the Hilbert space to any ``Seifert manifold with boundary,'' possessing a nondegenerate \(S^{1}\) fibration.  Necessarily, the boundary of such a manifold will be a disjoint union of two-tori.  Finally, the modular group acts geometrically on \(T^{2}\), so the refined Chern-Simons theory gives a unitary representation of  \(SL(2,\mathbb{Z})\) acting on \(\mathcal{H}_{T^{2}}\).

To fully specify the refined Chern-Simons theory, we must choose a simply-laced compact gauge group, \(G\), an integer level, \(k\in \mathbb{Z}\), and a continuous deformation parameter, \(\beta \in \mathbb{R}^{\geq 0}\).  It will be useful in the following formulas to define equivalent variables,
\begin{eqnarray}
q & = & \exp{\Big(\frac{2\pi i}{k+\beta y}\Big)} \\
t & = & \exp{\Big(\frac{2\pi i \beta}{k+\beta y}\Big)}
\end{eqnarray}
where \(y\) is the dual Coxeter number of \(G\).\footnote{The dual Coxeter number is given by \(y=N\) for \(SU(N)\), \(2N-2\) for \(SO(2N)\), 12 for \(E_{6}\), 18 for \(E_{7}\), and 30 for \(E_{8}\).}  Note that the unrefined limit corresponds to \(\beta \to 1\).

We start by describing the Hilbert space, \(\mathcal{H}_{T^{2}}\), associated to a \(T^{2}\) boundary.  Recall that in ordinary Chern-Simons theory, with gauge group, \(G\), and coupling, \(k\), the Hilbert space is given by the space of conformal blocks on \(T^{2}\) of the associated Wess-Zumino-Witten model.  This Hilbert space has a natural orthonormal basis, given by integrable representations of \(G\) at level \(k\).  In the Chern-Simons theory, this basis can be understood physically by taking the solid torus, \(D \times S^{1}\) and inserting a Wilson line in the representation \(\lambda\) running along \(\{0\} \times S^{1}\).  Then performing the path integral on this geometry gives the state \(\vert \lambda \rangle \in \mathcal{H}_{T^{2}}\).  

Note that to define this basis, we must decide which cycle of the \(T^{2}\) boundary will be filled in to make a solid torus.  This gives two natural bases associated with filling in the \(A\) and \(B\) cycles respectively.  The modular transformation, \(S \in SL(2,\mathbb{Z})\) then will be represented as a unitary operator that transforms the \(A\) basis into the \(B\) basis.

The vector space structure of \( \mathcal{H}_{T^{2}}\) does not change under refinement.  This is expected heuristically, since \(\beta\) is a continuous parameter that can be adiabatically changed from the unrefined theory \(\beta=1\) to the refined theory \(\beta \neq 1\).  This can be seen more precisely by noting that the metric, \(g_{i}\), defined below, vanishes for any representations that are not integrable at level \(k\).

However, under refinement the inner product on these spaces does change.  In ordinary Chern-Simons theory, the basis \(\vert \lambda \rangle\) is an orthonormal one so that,
\begin{equation}
\langle \lambda_{i} \vert \lambda_{j} \rangle = \delta_{ij}
\end{equation}
In the refined theory, this natural basis remains orthogonal but the normalization is nontrivial,
\begin{equation}
\langle \lambda_{i} \vert \lambda_{j} \rangle = g_{i}\delta_{ij}
\end{equation}
where \(g_{i}\) is the metric defined by taking the Macdonald inner product of the Macdonald polynomial, \(M_{\lambda_{i}}\) with itself (see Appendix \ref{app:mac} for background on Macdonald polynomials).  In the special case when \(\beta \in \mathbb{Z}^{>0}\), the metric factor is given explicitly by,
\begin{equation}
g_{i} \equiv \prod_{\alpha \in R_{+}}\prod_{m=0}^{\beta - 1}\frac{1-t^{\langle \rho, \alpha \rangle}q^{\langle \lambda_{i}, \alpha \rangle + m}}{1-t^{\langle \rho, \alpha \rangle}q^{\langle \lambda_{i}, \alpha \rangle - m}}
\end{equation}
where the product is over the positive roots, \(\alpha\), and \(\rho\) is the Weyl vector, \(\rho = \frac{1}{2}\sum_{\alpha>0}\alpha\).  For more general choices of \(\beta\), there exists a combinatorial formula for \(g_{i}\), which is given for the \(SO(2N)\) case in Appendix \ref{app:so2n}.  Note that we could rescale the \(\vert \lambda_{i} \rangle\) to obtain a normalized basis, but for subsequent formulas it is actually more convenient to leave the inner product nontrivial.

Now that we have described the Hilbert space associated to each \(T^{2}\) boundary, we should consider arbitrary Seifert three-manifolds, \(M\) with boundaries.  The refined Chern-Simons theory should assign a specific element of the Hilbert space, \(\Psi(M) \in \mathcal{H}_{T^{2}}^{\otimes n}\), to each three-manifold.

\begin{figure}[htp]
\centering
\includegraphics[scale=0.65]{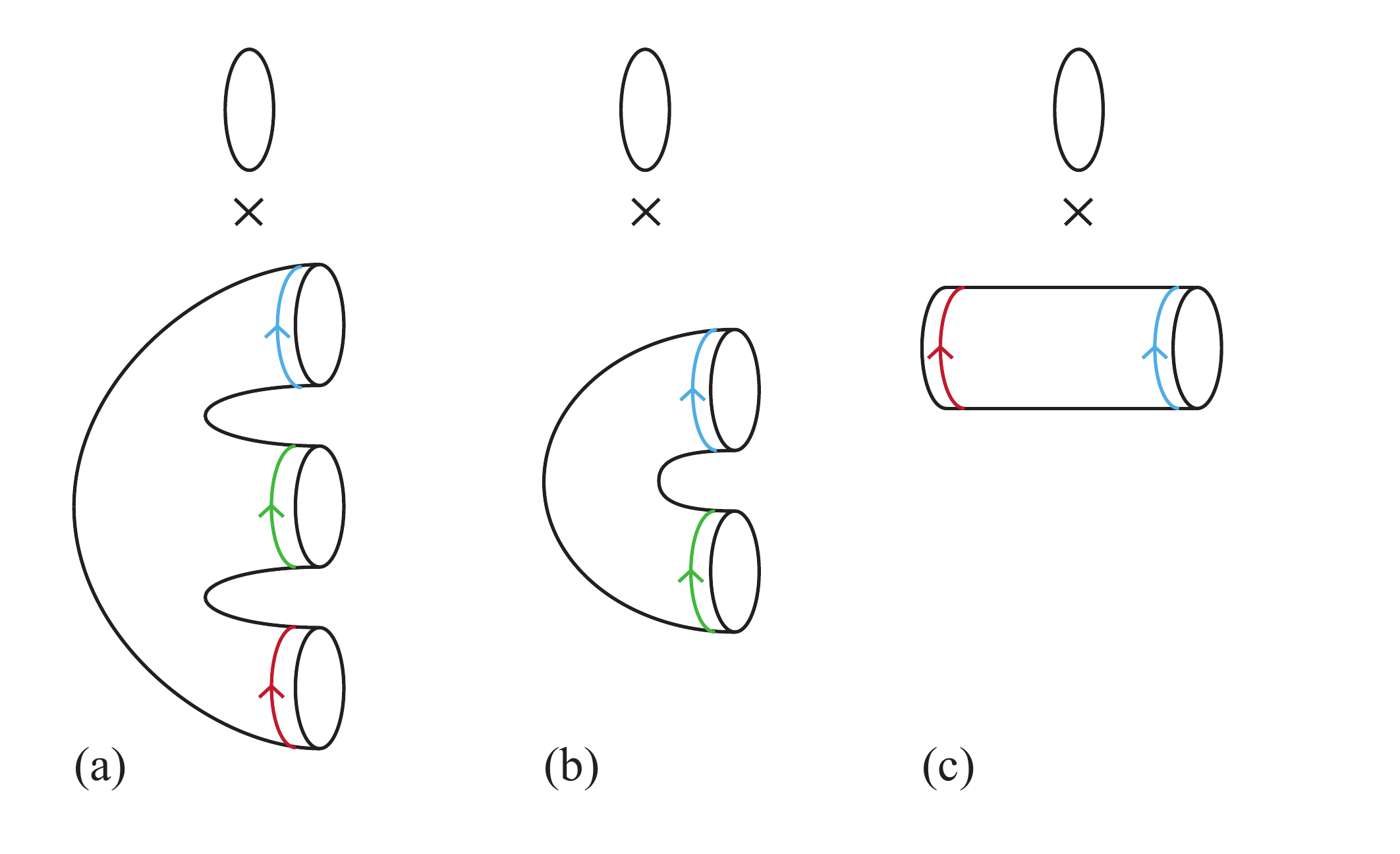}
\caption{The geometric building blocks for the refined Chern-Simons theory are shown.  The 3-punctured sphere times a circle, \(P \times S^{1}\), is shown in part (a), with the orientations of the three boundary \(T^{2}\)'s indicated by arrows.  The two propagators, which have the topology of \(\mathbb{C}^{*} \times S^{1}\) are shown in parts (b) and (c).  Note that they differ by the orientation of one of the boundary components.\label{fig:tqftdiagrams}}
\end{figure}

We begin by defining the theory on the ``pair of paints'' geometry given by the three-punctured sphere times a circle, \(P \times S^{1}\), shown in Figure \ref{fig:tqftdiagrams}(a).  In Figure \ref{fig:tqftdiagrams}, we have included Wilson lines wrapping the A-cycles of the \(T^{2}\) boundaries.  This depiction is useful for two reasons: first, we will use these Wilson lines to keep track of the orientation of the boundaries, since in a TQFT changing the orientation of a boundary takes the Hilbert Space \(\mathcal{H}\) to its dual, \(\mathcal{H}^{*}\).  

Second, recall from the discussion above that to specify a basis \(\vert \lambda_{i} \rangle\) of \(\mathcal{H}\), we must choose a cycle on \(T^{2}\).  In all of the following discussion, the element \(\vert \lambda_{i}\rangle\) will correspond to taking the \(T^{2}\) boundary and gluing in a solid torus, where the fiber \(S^{1}\) becomes contractible.  Then we insert a Wilson loop, in the representation \(\lambda_{i}\) running along one of the boundaries of the base pair of pants, \(P\) or annulus, \(\mathbb{C}^{*}\), and sitting at a point in the now-contractible \(S^{1}\) fiber.  Computing the path integral then gives the state \(\vert \lambda_{i} \rangle\).  We have depicted this choice of basis graphically by showing the Wilson loops on \(P\).

Then the wavefunction of the refined Chern-Simons theory on \(P \times S^{1}\) is given by, 
\begin{equation}
\Psi(P) = \sum_{i}\frac{1}{g_{i}S_{0i}}\langle \lambda_{i} \vert \langle \lambda_{i} \vert \langle \lambda_{i} \vert
\end{equation}
where \(g_{i}\) is the metric defined above and \(S_{0i}\) can be thought of as the \((q,t)\)-dimension of the representation \(\lambda_{i}\).  This \((q,t)\)-dimension is a refinement of the quantum-dimension that appears in ordinary Chern-Simons theory, and is given by,
\begin{equation}
S_{0i} = S_{00}\textrm{dim}_{q,t}(\lambda_{i}) \equiv S_{00}M_{\lambda_{i}}(t^{\rho}) = S_{00}\prod_{m=0}^{\beta-1}\prod_{\alpha>0}\frac{q^{\frac{\langle \lambda_{i}, \alpha\rangle + m}{2}}t^{\frac{\langle \rho, \alpha \rangle}{2}}- q^{-\frac{\langle \lambda_{i}, \alpha\rangle + m}{2}}t^{-\frac{\langle \rho, \alpha \rangle}{2}}}{q^{\frac{m}{2}}t^{\frac{\langle \rho, \alpha \rangle}{2}}- q^{-\frac{m}{2}}t^{-\frac{\langle \rho, \alpha \rangle}{2}}}
\end{equation}
where \(M_{\lambda_{i}}\) is the Macdonald polynomial for the gauge group \(G\)  and representation \(\lambda_{i}\), and \(S_{00}\) is a normalization factor defined below.

We must also define the theory on propagator geometries of the form \(\mathbb{C}^{*} \times S^{1}\) as shown in Figure \ref{fig:tqftdiagrams}(b) and \ref{fig:tqftdiagrams}(c).  The propagators differ by the orientation of one of the boundary components.  The first propagator, shown in Figure \ref{fig:tqftdiagrams}(b), is given by,
\begin{equation}
\Psi(\eta) = \sum_{i}\frac{1}{g_{i}}\langle \lambda_{i} \vert \langle \lambda_{i} \vert 
\end{equation}
while the second propagator, shown in Figure \ref{fig:tqftdiagrams}(c), is the same except for a reversal of orientation,
\begin{equation}
\Psi(\delta) = \sum_{i}\frac{1}{g_{i}}\vert \lambda_{i} \rangle \langle \lambda_{i} \vert 
\end{equation}
With this information, we can now compute the partition function of refined Chern-Simons theory on manifolds of the form \(\Sigma \times S^{1}\).   Remembering that \(\langle \lambda_{i} \vert \lambda_{i} \rangle = g_{i}\), we find for a genus g Riemann surface \(\Sigma\),
\begin{equation}
Z(\Sigma \times S^{1}) = \sum_{i}\frac{(g_{i})^{g-1}}{\big(S_{0i}\big)^{2g-2}}
\end{equation}

To obtain more complicated geometries, we must now understand how the modular group, \(SL(2,\mathbb{Z})\) acts on the Hilbert Space \(\mathcal{H}\).  Recall that the modular group can described by the generators, \(S\) and \(T\), subject to the relations,
\begin{equation}
S^{4} = 1 \qquad\qquad (ST)^{3} = S^{2}
\end{equation}
The refined Chern-Simons theory gives a representation of this group acting on \(\mathcal{H}\), which we can describe by computing the matrix elements of \(S\) and \(T\), provided they satisfy the above relations.\footnote{Because the inner product is nontrivial in these conventions, inserting a complete basis of states is given by, \(1 = \sum_{i}\frac{1}{g_{i}}\vert \lambda_{i}\rangle\langle \lambda_{i} \vert\).  To keep track of these additional \(g_{i}^{-1}\) factors when computing matrix elements, it is sometimes convenient to think of \(g_{i}\) as a lowering metric and \(g_{i}^{-1}\) as a raising metric, so that \(K_{i}^{\;\; j} = g_{j}^{-1}K_{ij}\).  We will use this notation below when we discuss knot computations.}  As before, these matrices are deformations of the Wess-Zumino-Witten \(S\) and \(T\) matrices for ordinary Chern-Simons.  The S-matrix is given by,
\begin{equation}
\langle \lambda_{i} \vert S \vert \lambda_{j} \rangle = S_{ij} \equiv S_{00}M_{\lambda_{i}}(t^{-\rho})M_{\lambda_{j}}(t^{-\rho}q^{-\lambda_{i}})
\end{equation}
where \(S_{00}\) is given by,
\begin{equation}
S_{0 0} = i^{\vert \Delta_{+}\vert}\vert P/Q \vert^{-1/2} (k+\beta y)^{-\frac{r}{2}}\prod_{m=0}^{\beta-1}\prod_{\alpha > 0}(q^{-m/2}t^{-(\alpha,\rho)/2}-q^{m/2}t^{(\alpha,\rho)/2}) \label{eqn:s00}
\end{equation}
where \(\vert \Delta_{+}\vert\) is the number of positive roots, \(y\) is the dual Coxeter number, and \(r\) is the rank of \(G\).  Here \(P\) is the weight lattice and \(Q\) is the root lattice, so that \(\vert P/Q \vert\) is the number of points in the fundamental cell of this quotient lattice.  The T matrix is given by,
\begin{equation}
\langle \lambda_{i} \vert T \vert \lambda_{j} \rangle = T_{i j} \equiv g_{i}q^{\frac{1}{2}(\lambda_{i},\lambda_{i})}t^{(\lambda_{i},\rho)}t^{\frac{\beta -1}{2}(\rho,\rho)}q^{-\frac{k}{2y}(\rho,\rho)} \delta_{ij}
\end{equation}

For the \(SU(N)\) case, Kirillov has proven in \cite{Kirillov1995q} that the \(S\) and \(T\) matrices satisfy the defining relations of \(SL(2,\mathbb{Z})\). In \cite{ChFourier2}, Cherednik generalized this result to arbitrary root systems by studying the \(SL(2,\mathbb{Z})\) action on the corresponding Double Affine Hecke Algebras.\footnote{In this paper we will only explicitly discuss the simply laced gauge groups, since these are simplest both mathematically and physically.  To obtain non-simply laced gauge groups, it should be possible to start with the simply laced \(ADE\) \((2,0)\) theory and introduce outer automorphism twists along the \(S^{1}\), as in \cite{Vafa:1997mh, Witten:2011zz}.  Although we will not study that construction here, the work of Cherednik provides further evidence that the refined Chern-Simons theory exists for general gauge groups.  It is interesting to note that in order for this \(SL(2,\mathbb{Z})\) representation to exist, and to ensure that the Hopf Link invariants are symmetric in the two representations, we must use the \emph{symmetric} rather than the ordinary Macdonald polynomials associated to the root system, \(R\).  These are the Macdonald polynomials used in Cherednik's work and were first defined by Macdonald under the name of \((R, R^{\vee})\) polynomials.  Of course, these polynomials agree with ordinary Macdonald polynomials for simply laced groups, but for non-simply laced groups, they are defined by using a modified inner product (see \cite{ChFourier} for details).}

As a simple application of this \(SL(2,\mathbb{Z})\) action, we can compute the amplitude on \(S^{3}\).  Using the Heegaard splitting of \(S^{3}\), its geometry is given by taking two solid tori (with no Wilson loops inserted), and gluing them after acting with the \(S\) operator.  This gives,
\begin{equation}
Z(S^{3}) = \langle 0 \vert S \vert 0 \rangle = S_{00}
\end{equation}

We have now given the full structure of the refined Chern-Simons theory as a restricted topological quantum field theory.  However, for computations it is helpful to explain another set of operators.  We define the operators \(\mathcal{O}_{i}\) by the property that,
\begin{equation}
\mathcal{O}_{i}\vert 0 \rangle = \vert \lambda_{i} \rangle
\end{equation}
Then we can ask what happens when we collide two of these operators.  The result should have the form,
\begin{equation}
\mathcal{O}_{i}\mathcal{O}_{j}\vert 0 \rangle = \sum_{k}N^{k}_{ij}\mathcal{O}_{k}\vert 0 \rangle
\end{equation}
In the unrefined case, this corresponds to placing two Wilson lines in representations \(\lambda_{i}\) and \(\lambda_{j}\) on top of each other.  Then the \(N^{k}_{ij}\) are simply the Littlewood-Richardson coefficients that arise from decomposing the tensor product, \(\lambda_{i} \otimes \lambda_{j}\).  In the refined case, these coefficients are deformed to the (q,t) Littlewood-Richardson coefficients, associated with decomposing the product of Macdonald polynomials (see Appendix \ref{app:mac}).  

Note that we could also understand these coefficients by taking the pant amplitude, \(P \times S^{1}\), but with Wilson loops wrapping the fiber \(S^{1}\) instead of the base.  This can be achieved by acting with the modular \(S\) matrix that exchanges the cycles of the \(T^{2}\) boundaries.  As explained in \cite{CSRefined}, this line of reasoning leads directly to the Verlinde formula for \(N^{k}_{ij}\).

\section{Refined Kauffman Invariants \label{sec:knots}}
Now that we have explained the structure of refined Chern-Simons theory as a TQFT, we can use it to compute new knot invariants.  In ordinary Chern-Simons theory, it is well known that if we introduce some Wilson loop in representation \(R\) along a knot \(\mathcal{K}\), then the expectation value of the Wilson loop gives a topological invariant of the knot.  

As explained above, we can also introduce Wilson loops in the refined Chern-Simons theory, but we are restricted to only choosing loops that are preserved by the \(U(1)\) action on the Seifert Manifold.  In this section we will focus on knots inside \(S^{3}\), so the \(U(1)\) condition restricts us to considering torus links.  To see this more explicitly, let us describe \(S^{3}\) as the locus in \(\mathbb{C}^{2}\), with coordinates \(z_{1}\), \(z_{2}\), where we require,
\begin{equation}
\vert z_{1} \vert^{2} + \vert z_{2} \vert^{2} = 1
\end{equation}
Here the \(S^{3}\) is naturally realized as a \(T^{2}\) fibration over the interval, where the coordinate on the interval is given by \(\vert z_{1}\vert^{2}\).  The \((1,0)\) cycle of the \(T^{2}\) comes from phase rotations of \(z_{1}\), while the \((0,1)\) cycle comes from phase rotations of \(z_{2}\).  Now we can consider the intersection of the \(S^{3}\) with the locus,
\begin{equation}
z_{1}^{n} = z_{2}^{m}
\end{equation}
The intersection is simply the \((n, m)\) torus link in \(S^{3}\).  We have not yet specified the \(U(1)\) action, but the most natural choice is \((z_{1}, z_{2}) \to (e^{i\theta m}z_{1}, e^{i\theta n}z_{2}) \) \footnote{  Note that we also could have chosen the action \((z_{1},z_{2}) \to (e^{i\theta m f}z_{1},e^{i\theta n f  }z_{2})\) for any \(f\in \mathbb{Z},\;\; f \neq 0\).  For any choice of \(f\), the \(U(1)_{f}\) action cannot be continuously deformed to the canonical \(f=1\) choice without passing through configurations that break supersymmetry, where the refined Chern-Simons theory is not defined.  Thus a priori, it is difficult to prove that the resulting Wilson loop expectation values are independent of the choice of \(f\).  However, in all the examples that we have checked the resulting expectation values do not depend on \(f\), up to trivial framing factors.}.  Recall that in the definition of Seifert manifolds, the \(U(1)\) action was only required to be semi-free, which is important here since every point on the circle \(z_{2}=0\) is fixed by the \(\mathbb{Z}_{m}\) subgroup, generated by \(e^{2\pi i/m}\).  

As explained above, the expectation value of the refined Chern-Simons theory in the presence of a torus knot can be computed using only the \(S\) and \(T\) matrices, and the metric, \(g_{i}\).  Above, we defined a knot operator that inserts a Wilson loop in the interior of the solid torus geometry, \(M_{L}\), so that,
\begin{equation}
\mathcal{O}^{(0, 1)}_{i}\vert 0 \rangle = \vert i \rangle
\end{equation}
Then the expectation value of the unknot is simply given by,
\begin{equation}
\langle 0 \vert \mathcal{O}^{(0, 1)}_{i} S \vert 0 \rangle = S_{i0}
\end{equation}
The operator that inserts the \((n, m)\) torus knot of interest is obtained by acting with an element of \(SL(2,\mathbb{Z})\) that maps \((0,1)\) to \((n,m)\),
\begin{equation}
K = \left( \begin{array}{cc}
a & n \\
b & m \\
\end{array} \right) \in SL(2,\mathbb{Z})
\end{equation}
Then the representation of \(K\) acting on the torus Hilbert Space can be written explicitly as a string of \(S\) and \(T\) matrices.  Then the \((n,m)\) operator is given by,
\begin{equation}
\mathcal{O}^{(n, m)} = K\mathcal{O}^{(0, 1)}K^{-1}
\end{equation}
In order to expand this out, we must use the \((q,t)\) Littlewood-Richardson coefficients, \(N^{k}_{\;\; ij}\),
\begin{equation}
\mathcal{O}^{(0, 1)}_{i}\vert j \rangle = \sum_{k}N^{k}_{\;\; ij}\vert k \rangle
\end{equation}
Putting these ingredients together we find an explicit formula for the \((n, m)\) torus knot invariant,
\begin{equation}
Z\big(T(n,m)\big) = \langle 0 \vert \mathcal{O}^{(n, m)}_{i}S\vert 0 \rangle = \langle 0 \vert K \mathcal{O}^{(0, 1)}_{i} K^{-1} S\vert 0 \rangle = \sum_{j,k,l} K_{0 k}N^{k}_{\;\; ij}(K^{-1})^{j}_{\;\; l}S^{l}_{\;\; 0}
\end{equation}
We will use this formula below to compute refined Chern-Simons \(SO(2N)\) knot invariants, but before doing so we clarify the relationship between refined Chern-Simons and other knot invariants.

\subsection{Relation to Knot Homology}

It is worth explaining here the relation of this approach to previous studies of knot homology in string theory \cite{Gukov:2004hz, Gukov:2005qp, Dunfield:2005si, Gukov:2007tf, Witten:2011zz, Gukov:2011ry}.  The starting point for the refined Chern-Simons theory comes from M-theory, where a stack of M5-branes wrapping the \(S^{3}\) intersects another stack of \(M5\) branes along the knot \(\mathcal{K}\).  If we kept the full space of BPS states at this intersection, then in accordance with the proposal of \cite{Gukov:2004hz}, this should describe the full knot homology associated to \(\mathcal{K}\).  Here the choice of the knot homology group (\(G = SU(N)\), \(SO(N)\), \(\cdots\)) corresponds to the insertion of \(N\) M5-branes wrapping \(S^{3}\), with the possible addition of orientifold planes for the orthogonal groups and the action of outer automorphisms for non-simply laced groups.  We denote this space by \(\mathcal{H}^{G}_{BPS}\).

This space of BPS states comes equipped with two natural gradings, \(S_{1}\) and \(S_{2}\), coming from the spins of the BPS states in \(\mathbb{C} \times \mathbb{C}\).  The Poincare polynomial in a given knot homology theory comes from computing the trace,
\begin{equation}
\mathcal{P}_{G}(\mathcal{K}) = \mathrm{Tr}_{\mathcal{H}^{G}_{BPS}}\mathbf{q}^{2(S_{1}-S_{2})}\mathbf{t}^{2S_{1}}
\end{equation}
where we have made the change of variables, \(\mathbf{q} = \sqrt{t}\) and \(\mathbf{t} = - \sqrt{q/t}\).  However, as emphasized in \cite{Witten:2011zz}, this trace cannot be computed as an index.  If we were to extend the trace to the entire M-theory Hilbert space, \(\mathcal{H}\), then non-BPS states would contribute.  Using only these gradings, the only genuine index that can be created is the euler characteristic,
\begin{equation}
P_{G}(\mathcal{K}) = \mathrm{Tr}_{\mathcal{H}^{G}_{BPS}}(-1)^{2S_{2}}\mathbf{q}^{2(S_{1}-S_{2})}
\end{equation}
which simply computes the ordinary quantum knot invariants coming from Chern-Simons theory.

The key to the refined Chern-Simons construction is the additional grading for torus knots that comes from the \(U(1)_{R}\) symmetry.  With this new grading it now becomes possible to compute a new index that contains refined information about the knot homology,
\begin{equation}
Z_{G}(\mathcal{K}) = \mathrm{Tr}_{\mathcal{H}^{G}_{BPS}}(-1)^{2S_{R}}\mathbf{q}^{2(S_{1}-S_{2})}\mathbf{t}^{2(S_{1}-S_{R})}
\end{equation}

Thus, the existence of the refined Chern-Simons theory makes the mathematical prediction that there should exist a new grading on the knot homology of torus knots.  A promising avenue for identifying this grading is the recent work of \cite{Rasmussen:talk, HilbertHomfly}, connecting the representation theory of Rational Double Affine Hecke Algebras (DAHA) to the HOMFLY and Khovanov-Rhozansky invariants of torus knots.  The DAHA plays a central role in the theory of Macdonald polynomials, so it is natural to suspect that their work may connect directly with refined Chern-Simons theory.  

Another method for computing the HOMFLY polynomial and Superpolynomial has been recently proposed by Cherednik \cite{2011CherednikJones}.  This method has the computational advantage that it directly uses the \(SL(2,\mathbb{Z})\) action on the DAHA, and does not require multiplying large matrices, as in the current refined Chern-Simons approach.  It would be interesting to understand Cherednik's method from a physics perspective.

In general, the index computed by the refined Chern-Simons theory will include negative signs, and will simply be different from the Poincare polynomial computed by Khovanov-Rozansky theory.  Surprisingly, however, it was found in \cite{CSRefined} that the large \(N\) behavior of Khovanov-Rozansky theory, encoded in the superpolynomial, could be reconstructed from the \(SU(N)\) refined Chern-Simons theory by using a special change of variables.  

For the \(SO(2N)\) case, a simple change of variables exists for the Hopf link (as in the \(SU(N)\) case), connecting with the Kauffman Homology of \cite{Gukov:2005qp, 2007KR}.  However, we find that no such change of variables exists for general torus knots, suggesting that the \(SO(2N)\) refined Chern-Simons theory computes genuinely new invariants for torus knots.

\subsection{Example: The Hopf Link}
Below we compute the \(SO(2N)\) refined Chern-Simons invariant for the Hopf link with both components colored by the fundamental representation, following a computational procedure similar to that used in \cite{CSRefined}.  Recall that the Hopf link knot invariant can be computed simply by evaluating an element of \(S\),
\begin{equation}
\overline{\textrm{Z}}(\textrm{Hopf},SO(2N)) = S_{V\, V}
\end{equation}
where \(V\) denotes the fundamental representation and the bar indicates that this is an unnormalized amplitude.  To simplify our results, it is helpful to normalize by the unknot amplitude,
\begin{equation}
\overline{\textrm{Z}}(\bigcirc,SO(2N)) = S_{\centerdot \, V} = \frac{t^{(2N-1)/2}-t^{-(2N-1)/2}}{t^{1/2}-t^{-1/2}}+1
\end{equation}
Normalizing by the unknot gives the general answer,
\begin{eqnarray}
\textrm{Z}(\textrm{Hopf}, SO(2N)) & = & \frac{\overline{\textrm{Z}}(\textrm{Hopf}, SO(2N))}{\overline{\textrm{Z}}(\bigcirc, SO(2N))} \\
& = & qt^{N-1}+1+q^{-1}t^{-(N-1)}+\frac{t^{(2N-3)/2}-t^{-(2N-3)/2}}{t^{1/2}-t^{-1/2}} \nonumber
\end{eqnarray}
Let us make the following change of variables,
\begin{eqnarray}
\mathbf{q} & = & \sqrt{t} \\
\mathbf{t} & = & -\sqrt{q/t} \nonumber \\
\mathbf{a} & = & t^{(2N-1)/2} \nonumber
\end{eqnarray}
Note that from the perspective of the large N dual, this is very natural since \(\mathbf{a}\) should correspond to the Kahler class of the resolved conifold.  This is the natural generalization of the \(SU(N)\) change of variables for the Hopf link used in \cite{CSRefined}.
Making this substitution, we obtain the superpolynomial,
\begin{equation}
Z(\textrm{Hopf}, SO(2N)) = \mathbf{t}^{2}\mathbf{q}\mathbf{a} + 1 +\mathbf{t}^{-2}\mathbf{q}^{-1}\mathbf{a}^{-1} +\frac{\mathbf{a}\mathbf{q}^{-2}-\mathbf{a}^{-1}\mathbf{q}^{2}}{\mathbf{q}-\mathbf{q}^{-1}}
\end{equation}

As a check of our methods, following Gukov and Walcher \cite{Gukov:2005qp} we use the fact that \(SO(4)\cong SU(2)\times SU(2)\) and the fundamental representation of \(SO(4)\) corresponds to the \((\mathbf{2},\mathbf{2})\) representation.  This implies that the SO(4) knot homology invariant should be equal to the square of the SU(2) Khovanov poincare polynomial, and we find perfect agreement.  Thus, for the Hopf Link, the refined Chern-Simons \(SO(2N)\) invariant agrees with the expected Kauffman Homology after a simple change of variables.

\subsection{Example: The Trefoil Knot}
We can follow a similar procedure to compute the refined \(SO(2N)\) invariant in the fundamental representation associated to the Trefoil, or \(T(2,3)\) knot.  As explained above, we must evaluate 
\begin{equation}
\langle 0 \vert \mathcal{O}^{(2,3)}_{V} S \vert 0 \rangle
\end{equation}
where the subscript \(V\) indicates that the knot is in the fundamental representation of \(SO(2N)\).  

\begin{figure}[htp]
\centering
\includegraphics[scale=0.18]{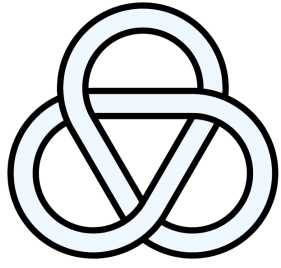}
\caption{The Trefoil Knot, \(T(2,3)\) \label{fig:m2mo5}}
\end{figure}
The result for the general normalized polynomial is,
\begin{eqnarray}
Z_{V}\Big(T(2,3), SO(2N)\Big) & = & \mathbf{t}^{6}\mathbf{q}^{2}\mathbf{a}^{4}+\mathbf{t}^{4}\mathbf{q}^{-2}\mathbf{a}^{4}+\mathbf{t}^{4}\mathbf{q}\mathbf{a}^{3}-\mathbf{t}^{4}\mathbf{q}^{-1}\mathbf{a}^{3}-\mathbf{t}^{4}\mathbf{q}^{2}\mathbf{a}^{2}+\mathbf{t}^{2}\mathbf{a}^{2} \nonumber \\
& & -\mathbf{t}^{2}\mathbf{q}^{-2}\mathbf{a}^{2}-\mathbf{t}^{2}\mathbf{q}\mathbf{a}+\mathbf{t}^{2}\mathbf{q}^{-1}\mathbf{a}
\end{eqnarray}

This refined Chern-Simons answer is structurally similar to the conjectured Kauffman homology result \cite{Gukov:2005qp} (as it must be, since they both reduce to the quantum \(SO(2N)\) invariant in the limit \(\mathbf{t}\to -1\)), but it can be seen straightforwardly that there does not exist a change of variables relating the two.

By the same method, we can also compute the normalized invariants associated to the spinor representation, \(S\).  For small values of \(N\) we find,
\begin{eqnarray}
Z_{S}\Big(T(2,3), SO(4)\Big) & = & (-\mathbf{q}\mathbf{t})^{-3}\big(\mathbf{t}^{6}\mathbf{q}^{10}+\mathbf{t}^{4}\mathbf{q}^{6}-\mathbf{t}^{4}\mathbf{q}^{4}\big) \nonumber \\
Z_{S}\Big(T(2,3), SO(6)\Big) & = & (-\mathbf{q}\mathbf{t})^{-3/2}\big(\mathbf{t}^{6}\mathbf{q}^{18}+\mathbf{t}^{4}\mathbf{q}^{14}-\mathbf{t}^{4}\mathbf{q}^{8}\big) \\
Z_{S}\Big(T(2,3), SO(8)\Big) & = & \mathbf{q}^{6} \mathbf{t}^{2} - \mathbf{q}^{8} \mathbf{t}^{2} - \mathbf{q}^{12} \mathbf{t}^{2} + \mathbf{q}^{14} \mathbf{t}^{2} - \mathbf{q}^{16} \mathbf{t}^{4} - \mathbf{q}^{20} \mathbf{t}^{4} + 
 \mathbf{q}^{22} \mathbf{t}^{4} + \mathbf{q}^{26} \mathbf{t}^{4} + \mathbf{q}^{30} \mathbf{t}^{6}\nonumber
\end{eqnarray}

\subsection{Example: The General $T(2,2m+1)$ Torus Knot}
By studying the invariants of the fundamental representation, \(V\), for the \(T(2,2m+1)\) torus knots, we find the following general formula for the refined Chern-Simons invariants,
\begin{eqnarray}
&& Z_{V}(T(2,2m+1),SO(2N)) = \mathbf{a}^{2}\mathbf{t}^{2} + \mathbf{a}^{2m+2}\mathbf{q}^{2m}\mathbf{t}^{4m+2}\Big(1-\mathbf{a}^{-2}\mathbf{t}^{-2}\Big)\Bigg(\sum_{i=0}^{m}\mathbf{q}^{-4i}\mathbf{t}^{-2i} \nonumber \\
& & +\sum_{j=1}^{2m-1}\mathbf{a}^{-j}\mathbf{q}^{-j}\mathbf{t}^{-2j}\Big(\sum_{i=1}^{\lceil(2m+1-j)/2\rceil} (\mathbf{q}^{4}\mathbf{t}^{2})^{-i+1} - \mathbf{q}^{-2}\sum_{i=1}^{\lfloor(2m+1-j)/2\rfloor}(\mathbf{q}^{4}\mathbf{t}^{2})^{-i+1} \nonumber\Big) \Bigg)
\end{eqnarray}

\subsection{Example: The $T(3,4)$ Knot}
Finally, we have studied the \(T(3,4)\) knot for invariants in the fundamental representation, \(V\).  For small gauge groups, we find,
\begin{eqnarray}
Z_{V}\Big(T(3,4), SO(4)\Big) & = & \mathbf{q}^{-24}\mathbf{t}^{-12}\Big(1 + \mathbf{q}^{4} \mathbf{t}^{2} - \mathbf{q}^{6} \mathbf{t}^{2} + \mathbf{q}^{6} \mathbf{t}^{4} - \mathbf{q}^{10} \mathbf{t}^{4}\Big)^{2} \\
Z_{V}\Big(T(3,4), SO(6)\Big) & = & \mathbf{q}^{-36}\mathbf{t}^{-12}\Big(1 + \mathbf{q}^{4} \mathbf{t}^{2} + \mathbf{q}^{6} \mathbf{t}^{2} - \mathbf{q}^{8} \mathbf{t}^{2} - \mathbf{q}^{10} \mathbf{t}^{2} + \mathbf{q}^{6} \mathbf{t}^{4} + 2 \mathbf{q}^{8} \mathbf{t}^{4} - 
 \mathbf{q}^{12} \mathbf{t}^{4} \nonumber \\
 & & - 2 \mathbf{q}^{14} \mathbf{t}^{4} - \mathbf{q}^{16} \mathbf{t}^{4} + \mathbf{q}^{18} \mathbf{t}^{4} + 2 \mathbf{q}^{12} \mathbf{t}^{6} + 
 \mathbf{q}^{14} \mathbf{t}^{6} - 3 \mathbf{q}^{16} \mathbf{t}^{6} - 2 \mathbf{q}^{18} \mathbf{t}^{6} \nonumber \\
 & & + \mathbf{q}^{22} \mathbf{t}^{6} + \mathbf{q}^{24} \mathbf{t}^{6} +
 \mathbf{q}^{16} \mathbf{t}^{8} - 2 \mathbf{q}^{22} \mathbf{t}^{8} - \mathbf{q}^{24} \mathbf{t}^{8} + 2 \mathbf{q}^{26} \mathbf{t}^{8}\Big)
\end{eqnarray}

\section{The Large N Limit \label{sec:largen}}
In this section we study the large N limit of \(SO(2N)\) refined Chern-Simons theory on \(S^{3}\).  Recall that the large N limit of ordinary \(SU(N)\) Chern-Simons theory on \(S^{3}\) is given by closed topological string theory on the resolved conifold \cite{Gopakumar:1998ki}.  The parameters on each side are related by,
\begin{eqnarray}
g_{s\; closed} & = & \frac{2\pi i}{k+N} \\
t  = Ng_{s\; closed} & = & \frac{2\pi i N}{k+N} \nonumber
\end{eqnarray}
where \(t\) is the Kahler parameter of the base \(\mathbb{P}^{1}\) in the resolved conifold.  Since \(SU(N)\) Chern-Simons theory is equivalent to open topological string theory on \(T^{*}S^{3}\) with N A-branes wrapping the \(S^{3}\), this is an example of a topological open-closed string duality analogous to the celebrated AdS/CFT correspondence.  This interpretation is reinforced by the observation that \(t = Ng_{s}\) takes the form of the usual `t Hooft parameter, since from the Chern Simons action, \(g_{s} = \frac{2\pi i}{k+N} = g_{open} ^{2}\).   The corresponding unrefined duality for \(SO(N)\) Chern Simons theory has also been studied in \cite{SinhaVafa}, where it was interpreted as an open-closed topological string duality in the presence of orientifolds.

In studying the refined \(SU(N)\) Chern-Simons theory, it was shown that a similar duality exists between \(SU(N)\) refined Chern-Simons and the refined topological string on the resolved conifold \cite{CSRefined}.  Here, we study this refined geometric transition in the presence of orientifolds.  We find that the large N limit of \(SO(2N)\) refined Chern-Simons is dual to refined, closed topological string theory on the resolved conifold, \(\mathcal{O}(-1)\oplus \mathcal{O}(-1) \to \mathbb{P}^{1}\) in the presence of an orientifold, \(I'\), that acts freely.

To describe the action of \(I'\), recall that in the linear sigma model description of the resolved conifold, we take four coordinates, \(X_{i}\) with charges \((1,\, 1,\, -1,\, -1)\) under a \(U(1)\) action.  Then the resolved conifold is given by,
\begin{equation}
\{\lvert X_{1} \rvert^{2} + \lvert X_{2} \rvert^{2} - \lvert X_{3} \rvert^{2} - \lvert X_{4} \rvert^{2} = r\}/U(1)
\end{equation}
The involution, \(I'\) acts by,
\begin{equation}
I' : \; (X_{1},X_{2},X_{3},X_{4}) \to(\overline{X}_{2},-\overline{X}_{1},\overline{X}_{4},-\overline{X}_{3})
\end{equation}
Note that \(I'\) acts freely on X, so that in \(X/I'\) the base \(\mathbb{CP}^{1}\) becomes \(\mathbb{RP}^{1}\).  

To see the duality explicitly, we take the partition function of $SO(2N)$ refined Chern-Simons theory on \(S^{3}\),
\begin{equation}
Z=S_{0 0} = \frac{1}{2\big(k+\beta (2N-2)\big)^{\frac{N}{2}}}\prod_{m=0}^{\beta-1}\prod_{\alpha > 0}(q^{-m/2}t^{-(\alpha,\rho)/2}-q^{m/2}t^{(\alpha,\rho)/2})
\end{equation}
We are interested in the free energy, \(F = -\log{Z}\) and will only keep the factors that have non-trivial \(q\) and \(t\) dependence,
\begin{equation}
F = \ldots - \sum_{m=0}^{\beta-1}\sum_{\alpha>0}\log{\big(1-q^{m}t^{(\alpha,\rho)}\big)}
\end{equation}
Now using the properties of the D root system, this can be rewritten as,
\begin{equation}
F = -\sum_{m=0}^{\beta-1}\sum_{k=1}^{2N-1}f(k)\log{\big(1-q^{m}t^{k}\big)}
\end{equation}
where \(f(k)\) is given by,
\begin{equation}
f(k) = \left\{ \begin{array}{ll}
\frac{2N+1-k}{2} & \textrm{if $k<N$, $k$ odd}\\
\frac{2N-1-k}{2} & \textrm{if $k\geq N$, $k$ odd}\\
\frac{2N-k}{2} & \textrm{if $k<N$, $k$ even}\\
\frac{2N-2-k}{2} & \textrm{if $k\geq N$, $k$ even}
\end{array} \right.
\end{equation}
After some algebraic manipulation, the free energy can then be written as,
\begin{eqnarray}
& = & \sum_{d=1}^{\infty}\frac{1}{2d}\frac{t^{(2N-1)d}t^{d/2}q^{-d/2}}{(q^{d/2}-q^{-d/2})(t^{d/2}-t^{-d/2})}+\sum_{d=1}^{\infty}\frac{1}{d}\frac{t^{(N-\frac{1}{2})d}t^{d/2}q^{-d/2}}{(q^{d/2}-q^{-d/2})} \nonumber\\
& & -\sum_{d=1}^{\infty}\frac{1}{2d}\frac{t^{(2N-1)d}t^{d/2}q^{-d/2}}{(q^{d/2}-q^{-d/2})(t^{d/2}+t^{-d/2})}
\end{eqnarray}
Note that this looks like a refined version of the unoriented first-quantized structure seen in Equation \ref{eqn:forient}.  It would be interesting to understand the presence of plus signs in the denominator from the first-quantized perspective, in terms of a graviphoton background.

However, recall that in Section \ref{sec:ofolds}, we gave a definition of refinement in terms of computing a second-quantized M-theory trace.  For this interpretation, it is more natural to write the free energy as,
\begin{eqnarray}
F & = & \sum_{d=1}^{\infty}\frac{1}{d}\frac{t^{(2N-1)d}q^{-d/2}}{(q^{d/2}-q^{-d/2})(t^{d}-t^{-d})}+\sum_{d=1}^{\infty}\frac{1}{d}\frac{t^{(N-\frac{1}{2})d}t^{d/2}q^{-d/2}}{(q^{d/2}-q^{-d/2})}
\end{eqnarray}
Note that this expression precisely satisfies the integrality properties of Section \ref{sec:integrality} for refined topological string theory in the presence of orientifolds.  We can identify the Kahler class of the resolved conifold as,
\begin{equation}
Q = t^{2N-1}
\end{equation}
and rewrite the free energy as,
\begin{eqnarray}
F & = & \sum_{d=1}^{\infty}\frac{1}{d}\frac{Q^{d}q^{-d/2}}{(q^{d/2}-q^{-d/2})(t^{d}-t^{-d})}+\sum_{d=1}^{\infty}\frac{1}{d}\frac{Q^{d/2}t^{d/2}q^{-d/2}}{(q^{d/2}-q^{-d/2})}
\end{eqnarray}
Then the first term comes from oriented \(M2\) branes that move freely in the four noncompact dimensions, and the second term comes from unoriented \(M2\) branes wrapping \(\mathbb{RP}^{2}\).  The \(q\) and \(t\) dependent shifts in the numerator arise because the presence of the orientifold breaks \(SU(2)_{L} \times SU(2)_{R}\) down to \(U(1)_{1}\times U(1)_{2}\).  For this reason, as discussed in Section \ref{sec:ofolds}, the BPS contributions do not have to appear in full spin multiplets.  

Altogether, we have found a refined geometric transition for unoriented strings, given by,
\begin{equation}
Z_{open\; ref}(T^{*}S^{3}/I;q,t,N) = Z_{closed\; ref}(X/I' ;q,t)
\end{equation}
This transition has also given us a nontrivial test of the conjectured integrality properties of the refined topological string in the presence of orientifolds.  It would be interesting to use these refined geometric transitions to compute refined amplitudes for more complicated geometries \cite{p2progress}, such as the orientifold of local \(\mathbb{P}^{2}\) considered in \cite{Bouchard:2004iu}.

\acknowledgments{We thank Chris Beem, Ivan Cherednik, Tudor Dimofte, Ori Ganor, Eugene Gorsky, Sergei Gukov, Daniel Krefl, and Jacob Rasmussen for helpful discussions.  We also thank Marcos Mari\~{n}o for comments on the manuscript.  K.S. thanks the Kavli Institute for Theoretical Physics (research supported in part by DARPA under Grant No. HR0011-09-1-0015 and by the National Science Foundation under Grant No. PHY05-51164) for warm hospitality while this work was being done. The authors also thank the Simons Workshop in Mathematics and Physics 2011 for providing a stimulating work environment.  The research of K.S. is supported by the Berkeley Center for Theoretical Physics and the Lawrence Berkeley National Laboratory.  The research of M.A. is supported in part by the Berkeley Center for Theoretical Physics, by the National Science Foundation (award number 0855653), by the Institute for the Physics and Mathematics of the Universe, and by the US Department of Energy under Contract DE-AC02-05CH11231.
}

\appendix

\section{The $(2,0)$ Theory on Seifert Three-Manifolds \label{sec:twist}}

In this appendix we explain in more detail how the refined Chern-Simons theory arises from wrapping \(M5\) branes on Seifert three-manifolds.

We begin by considering the \((2,0)\) theory on three-manifolds of the special form, \(M = \Sigma \times S^{1}\), where $\Sigma$ is a Riemann surface, and we take the rest of the worldvolume directions to be \(\mathbb{R}^{2}\times S^{1}_{\beta}\).  Since $\Sigma$ is curved the theory will be partially topologically twisted along $\Sigma$.

We can either work in M-theory, where the partial topological twist is implemented by the geometry, \(M \subset T^{*}M \), or in the \((2,0)\) theory by performing the topological twist directly.  For maximal generality, we will consider the \(A\), \(D\), and \(E\)-type \((2,0)\) theories together.  Recall that the \(D\)-type \((2,0)\) theory arises from placing \(M5\) branes on an orbifold, as we have considered in the body of the paper.  An \(M5\)-brane construction of the \(E\)-type theory is not known, but we can still describe it in terms of the \((2,0)\) theory.  In the following, we denote the choice of gauge group by \(G\).

The construction is simplified by working instead with five-dimensional \(\mathcal{N}=2\) Super-Yang-Mills theory, obtained by reducing on the trivial \(S^{1}\).  The \(\mathcal{N}=2\) supersymmetry algebra in five dimensions contains the maximal bosonic subalgebra, \(SO(4,1)_{E} \times SO(5)_{R}\).  Under the \(SO(4,1)_{E}\) rotation group and the \(SO(5)_{R}\)-symmetry, the supercharges transform as \((\mathbf{4},\mathbf{4})\). 

M5 branes wrapping \(S^{1}_{\beta} \times \mathbb{C} \times \Sigma \times S^{1} \subset S^{1}_{\beta} \times \mathbb{C}^{2} \times T^{*}(\Sigma \times S^{1})\) correspond to Yang-Mills on \(S^{1}_{\beta} \times \mathbb{R}^{2} \times \Sigma\) , so it is natural to study the subgroups \(SO(2)_{S_{1}} \times SO(2)_{\Sigma}\subset SO(4,1)_{E}\) and \(SO(2)_{R} \times SO(2)_{S_{2}} \subset SO(5)_{R}\).  Geometrically, \(SO(2)_{S_{1}}\) corresponds to rotations along the brane in \(\mathbb{R}^{2}\), \(SO(2)_{\Sigma}\) rotates the Riemann surface, \(SO(2)_{R}\) rotates the directions in the Calabi-Yau transverse to the brane, and \(SO(2)_{S_{2}}\) corresponds to rotations in the two noncompact transverse directions.\footnote{The convention of using \(S_{2}\) to denote part of the R-symmetry may seem strange from a field theory perspective, but from the brane picture it is quite natural since this R-symmetry rotates the the transverse noncompact directions.  This notation is especially natural when the \(M5\) branes are surface operators in a non-trivial geometrically engineered five-dimensional theory.}  The sixteen supercharges have the quantum numbers, \((\pm \frac{1}{2}, \pm\frac{1}{2}; \pm\frac{1}{2},\pm\frac{1}{2})\) under this subgroup.

The partial topological twist comes from taking the diagonal combination \(SO(2)'_{\Sigma} = (SO(2)_{\Sigma} \times SO(2)_{R})_{diag}\).  It is important to note that the groups involved in the twist are abelian; for this reason the supercharges will simultaneously have well-defined quantum numbers under both \(SO(2)'_{\Sigma}\) and \(SO(2)_{R}\).
Upon performing the topological twist, we keep only those supercharges that are neutral under 
\(SO(2)'_{\Sigma}\) so we are left with eight supercharges whose quantum numbers under \(SO(2)_{S_{1}} \times SO(2)_{R} \times SO(2)_{S_{2}}\) are, \((\pm \frac{1}{2}, \pm \frac{1}{2}, \pm \frac{1}{2})\) (see Table \ref{superchargetab4}).  These supercharges form an \(\mathcal{N}=4\) superalgebra in three dimensions.  \(\mathcal{N}=4\) supersymmetry in three dimensions has an \(SU(2)_{R} \times SU(2)_{S_{2}}\) R-symmetry, and the above \(SO(2)_{R}\times SO(2)_{S_{2}}\) symmetries are simply the respective abelian subgroups of the full R-symmetry.

\begin{table}[htp]
\centering
\begin{tabular}[c]{| c | | c | c | c | c |}
\hline
& $2S_1$ & $2S_2$ & $2S_R$ & $S_R-S_2$ \\
\hline
$Q^{11}_{+}$ & $+1$ & $+1$ & $+1$ & $0$ \\
\hline
$Q^{11}_{-}$ & $-1$ & $+1$ & $+1$ & $0$ \\
\hline
$Q^{22}_{+}$ & $+1$ & $-1$ & $-1$ & $0$ \\
\hline
$Q^{22}_{-}$ & $-1$ & $-1$ & $-1$ & $0$ \\
\hline
$Q^{21}_{+}$ & $+1$ & $-1$ & $+1$ & $+1$ \\
\hline
$Q^{21}_{-}$ & $-1$ & $-1$ & $+1$ & $+1$ \\
\hline
$Q^{12}_{+}$ & $+1$ & $+1$ & $-1$ & $-1$ \\
\hline
$Q^{12}_{-}$ & $-1$ & $+1$ & $-1$ & $-1$ \\
\hline
\end{tabular}
\caption{Supercharge Quantum Numbers in Three-Dimensional \(\mathcal{N}=4\) Supersymmetry.  The addition of the supersymmetric Chern-Simons term breaks this to \(\mathcal{N}=2\) supersymmetry, preserving only the supercharges that are neutral under \(S_{R}-S_{2}\).  \label{superchargetab4}}
\end{table}

Now recall that the refined Chern-Simons theory is defined by computing the index in the three-dimensional theory given by,
\begin{equation}
Z_{ref\;open}(X;q,t) = \textrm{Tr}\; (-1)^{2S_{1}}q^{S_{1}-S_{R}}t^{S_{R}-S_{2}}e^{-\beta H}
\end{equation}
Note that this index counts states annihilated by both \(Q^{11}_{+}\) and \(Q^{22}_{-}\), as can be seen by reading off the quantum numbers in Table \ref{superchargetab4}.  Here we have given a three-dimensional field-theoretic interpretation to the index, but we would also like to match up these symmetries with those in the original M-theory definition of the index in \cite{CSRefined}.  

From the geometric picture given above, it should be clear that the \(S_{1}\) and \(S_{2}\) symmetries agree with those defined in \cite{CSRefined}.  Originally the \(S_{R}\) symmetry was identified with a rotation in the fibers of \(T^{*}M\) that is transverse to the \(U(1)\) isometry of the Seifert manifold, M.  In our case, this Seifert manifold isometry is given by simply rotating the \(S^{1}\) in \(M=\Sigma \times S^{1}\).  This means that the \(S_{R}\) symmetry must rotate the cotangent fiber over \(\Sigma\), but this is precisely the \(U(1)_{R}\) rotation that appeared above in our topological twist.  Thus, we have given a purely field-theoretic identification of the symmetries involved in the refined Chern-Simons index.

Now we consider the more general case of M5 branes on a Seifert three manifold obtained by fibering the $S^1$ non-trivially over $\Sigma$. In this case, half of the supersymmetry of the theory is broken: after the partial topological twist, the theory on \(\mathbb{R}^{2,1}\) has only ${\cal N}=2$ supersymmetry in three dimensions.  This also corresponds to the fact that in this case \(T^{*}M\) is an honest Calabi-Yau three-fold (its holonomy group is precisely \(SU(3)\) and not a subgroup).  Recall that \(\mathcal{N}=2\) supersymmetry has a \(U(1)\) R-symmetry, but if \(M\) is an arbitrary three-manifold no additional symmetries will be present in the problem.  Of the symmetries described above, only the rotation symmetry \(S_{1}\) and the transverse R-symmetry, \(S_{2}\), will survive.  However, the key is that for the special case of a Seifert manifold, the \(S_{R}\) symmetry will also be preserved by the breaking from \(\mathcal{N}=4\) to \(\mathcal{N}=2\).  In this case all the supercharges are uncharged under \(S_{2}-S_{R}\), which becomes a new global symmetry.

One way to argue for this is to use the geometric description of the \(U(1)_{R}\) symmetry when $M$ is a Seifert three-manifold.  The argument presented in \cite{CSRefined} is based on the fact that one can use a nowhere vanishing vector field  $V$ rotating the $S^1$ fiber of the Seifert three-manifold to define, at each point on $\Sigma$ a two-plane in the fiber $T^*M$ -- this two plane is co-normal to $V$ in the natural symplectic structure on $T^*M$.  

However, we would also like to see directly that this symmetry survives in the partial twisting of the \((2,0)\) theory.  We start by asking about the effect of fibering the $S^1$ over $\Sigma$ in the $(2,0)$ theory and in the dimensionally reduced five-dimensional Yang-Mills theory -- where we reduce on the $S^1$ fiber of the Seifert three-manifold, as we did above. The answer is that for an $S^1$ bundle of degree \(p\) fibered over $\Sigma$, we obtain an ${\cal N}=2$ Chern-Simons coupling on $R^{2,1}$,
\begin{equation}
p \int_{\mathbb{R}^{2,1}}d^{2} \theta \textrm{ Tr}(\Sigma \mathcal{V}) \label{eqn:cssusy}
 \end{equation}
 where \(\mathcal{V}\) is the $\mathcal{N}=2$ vector multiplet, and $ \Sigma = \epsilon^{\alpha \beta} {\bar D}_{\alpha} D_{\beta}\mathcal{V}$ is the linear superfield.\footnote{Recently, in \cite{Cecotti:2011iy} it was argued that when wrapping \(M5\) branes on a three-manifold, Chern-Simons terms should arise from torsion in the first homology of \(M\).  Specifically, they argued that the factor \(\mathbb{Z}_{p}\in H_{1}(M,\mathbb{Z})\) should correspond to a Chern-Simons term at level \(p\).  This is perfectly consistent with our result since the first homology of a degree \(p\) circle bundle over a genus \(g\) Riemann surface, \(\Sigma\), is given by \(H_{1}(M,\mathbb{Z}) = \mathbb{Z}^{2g}\oplus \mathbb{Z}_{p}\).}  The crucial point is that this coupling is neutral under both $U(1)_r$ and $U(1)_R$, so turning it on does not break either symmetry.  This can be seen by expanding out the Chern-Simons term in components,
\begin{equation}
\label{eqn:csterm}
p \int_{\mathbb{R}^{2,1}}d^{3}x d^{2}\theta \textrm{ Tr}( \Sigma \mathcal{V}) = p\int_{\mathbb{R}^{2,1}}d^{3}x \mathrm{  Tr}\Big(A\wedge dA + \frac{2}{3}A\wedge A\wedge A+ i\overline{\chi}\chi - 2D\sigma\Big)
\end{equation}
For example, since \(A_{\mu}\) is uncharged under both \(U(1)_{r}\) and \(U(1)_{R}\), the Chern-Simons term does not break these symmetries. 

The origin of the Chern-Simons coupling is purely topological. The fastest way to see this is to take the circle fiber of \(M\) as the M-theory circle, and reduce the six dimensional \(M5\) brane theory to the five dimensional Yang-Mills theory living on a \(D4\) brane.  Now recall that the \(D4\) brane action includes a coupling,
\begin{equation}
\int F_{RR} \wedge \textrm{Tr}\Big(A \wedge dA + \frac{2}{3} A\wedge A \wedge A\Big) \label{eqn:wzcoup}
\end{equation}
Using the relationship between IIA and M-theory, $F_{RR}$ is simply the curvature of the M-theory circle bundle.  In our present case, we have
$$
\int_{\Sigma} F_{RR} = p
$$
so we obtain the expected Chern-Simons term. The rest of Equation \ref{eqn:csterm} is fixed by ${\cal N}=2$ supersymmetry. 

It may seem that we are using some very particular facts about the couplings on the D4 branes, but this is not the case.  The term in Equation \ref{eqn:wzcoup} really appears because the $(2,0)$ theory has a propagating self-dual two-form tensor. In any attempt to write down the action for such a theory, there is a peculiar Wess-Zumino type term that arises -- albeit involving the metric on the six-dimensional worldvolume \cite{Schwarz:1997mc, Aganagic:1997zq}. Using dimensional reduction to get the five dimensional Yang-Mills theory, the term \ref{eqn:wzcoup} arises with \(F_{RR}\) as the curvature of the circle bundle that we reduced on.  This more general argument also makes it clear that the Chern-Simons term appears for the \(E\)-type (2,0) theory even though it does not have a known M-theoretic brane construction.

Thus, in the specific case when $M$ is a Seifert three manifold, the theory has, an R-symmetry given by \(S_{2}\) and an additional $U(1)$ flavor symmetry generated by $S_{2}-S_R$. Therefore, when $M$ is a Seifert three manifold we can define the refined index that computes refined Chern-Simons theory.

It may seem that by using the dimensional reduction on the \(S^{1}\) fiber, the above topological twist is different from the usual geometric one obtained by wrapping a brane on \(M \subset T^{*}M\).  To see that they are the same, we would need to perform the three-dimensional partial topological twist directly in the \((2,0)\) theory.  This is rather difficult, but we can instead reduce the entire theory on a trivial circle and then study the twisting of the \(D4\)-brane theory on \(M\).  

This gives two different ways of reducing the theory.  First, we can use the above construction to give an \(\mathcal{N}=2\) theory in three dimensions, and then further reduce to an \(\mathcal{N}=(2,2)\) two-dimensional field theory.  Alternatively, we can reduce on the trivial circle first, and then perform the standard topological twist of the \(D4\) brane theory on the three-manifold, \(M\), leaving us with a two-dimensional field theory.  These procedures should agree, giving us a consistency check that the above twist is really the same as the standard twist.

Starting with the first approach, the dimensional reduction of the supersymmetric Chern-Simons term in Equation \ref{eqn:cssusy} is given by an \(\mathcal{N}=(2,2)\) twisted superpotential, 
\begin{equation}
\widetilde{W} = p\int d\theta^{+}d\overline{\theta}^{-} \Sigma^{2} = p\Bigg(4\sqrt{2}H\textrm{Re}(\sigma) + 4\sqrt{2}F_{01}\textrm{Im}(\sigma) + 4\textrm{tr}\Big(\overline{\lambda}_{+}\lambda_{-} + \lambda_{+}\overline{\lambda}_{-}\Big)\Bigg)
\end{equation}
where \(\Sigma\) is the two-dimensional super-field strength given by,
\begin{equation}
\Sigma = \sigma + i \sqrt{2}\Big(\theta^{+}\overline{\lambda}_{+} - \overline{\theta}^{-}\lambda_{-}\Big) + \sqrt{2}\theta^{+}\overline{\theta}^{-}\Big(H-iF_{01}\Big)
\end{equation}
where \(\sigma\) is a complex scalar whose real part comes from the three-dimensional real scalar and whose imaginary part comes from the Wilson line of the three-dimensional gauge field, \(\oint_{S^{1}_{\beta}}A\).
Upon integrating out the auxillary field \(H\) and properly taking \(F_{01}\) into account, we obtain the potential terms,
\begin{equation}
U = g^{2}p^{2}(\textrm{Re}(\sigma))^{2}+g^{2}p^{2}(\textrm{Im}(\sigma))^{2} \label{eqn:sigmapot}
\end{equation}
where \(g^{2}\) is the coupling constant coming from the kinetic term.

Now we would like to understand how these quadratic terms are reproduced by the partial topological twist of D4 branes wrapping \(M\).  More generally, we must understand what is special about Seifert manifolds from this perspective.  For example, a naive group theory analysis of the topological twist would not detect the additional symmetry that appears when \(M\) is a Seifert manifold.

Both of these problems can be resolved by making full use of the structure of \(M\) as a Seifert manifold.  By definition this means that \(M\) has a nowhere-vanishing vector field, \(v\), that acts as an isometry on \(M\).  By using the metric, this also implies the existence of a nowhere-vanishing one-form on \(M\), which we denote by \(\kappa\).

Upon choosing \(v\), the structure group of the tangent bundle, \(TM\), is reduced from \(SO(3)\) to \(SO(2)\), since the transition functions must respect the globally defined vector field, \(v\).  Because the structure group is reduced, the spin bundle over \(M\) will also split into two line bundles corresponding to the one-dimensional irreducible representations of \(SO(2)\) (see \cite{Nicolaescu1996, Mrowka1996} for a discussion of this splitting in the context of three-dimensional Seiberg-Witten Floer theory).  Finally, the same arguments imply that the cotangent bundle, \(T^{*}M\) will also split.  This splitting is crucial since it explains why the fields living on the brane have well-defined quantum numbers under the geometric symmetry, \(S_{R}\).

The topological twist coming from a brane wrapping \(M\subset T^{*}M\) means that in addition to the gauge fields, \(A_{\mu}\), the three scalars fields corresponding to fluctuations of the brane in the fiber directions of \(T^{*}M\) will also now transform as a one form.  Now by using the splitting of the cotangent bundle, any one-form on \(M\) can be decomposed as,
\begin{equation}
A = \phi\kappa + A_{\Sigma}
\end{equation}
where \(A_{\Sigma}\) is a one-form orthogonal to \(\kappa\).  

If we focus on the topologically twisted scalar fields, this means that we obtain one scalar and two vector components instead of the three vector components that we would get by topologically twisting on an arbitrary three-manifold.  Note that this looks like the field content that we would expect from performing a topological twist only on the base two-manifold, \(\Sigma\).  The difference is that here we have a globally nontrivial circle fibration, which is responsible for the additional mass terms that break half of the supersymmetry.

Now we want to see how these twisted superpotential terms of equation \ref{eqn:sigmapot} arise.  By using the above decomposition of \(A\), the gauge field kinetic term includes,
\begin{equation}
\int_{\mathbb{R}^{1,1}\times M} d(\phi\kappa)\wedge \star d(\phi\kappa)
\end{equation}
But as explained in \cite{Beasley:2005vf, Blau:2006gh, Kallen:2011ny}, \(d\kappa = p (\star\kappa)\) and \(\int \kappa\wedge d\kappa = p\), where \(p\) again is the Euler class of the circle bundle of \(M\).\footnote{One way to see this is to recall that, as argued in \cite{Beasley:2005vf, Blau:2006gh, Kallen:2011ny}, \(\int\kappa \wedge d\kappa = p\).  Further, if \(\kappa\) is normalized so that \(\langle \kappa, \kappa\rangle =1\) then we have, \(\int \kappa \wedge \star\kappa = \int \langle \kappa, \kappa \rangle d\mu = V\) where \(V\) is the volume of the three-manifold.  Finally, we know that \(d\kappa = f(\star\kappa)\) where \(f\) is some arbitrary function.  By choosing \(\kappa\) appropriately, \(f\) becomes a constant, and from the above facts we can deduce that \(d\kappa = (p/V)(\star \kappa)\).  We have dropped the volume dependence above, since it can be absorbed in the overall coupling constant which we have also omitted for simplicity.}  Thus we obtain the term,
\begin{equation}
\int_{\mathbb{R}^{1,1}}\phi^{2}\int_{M} d\kappa\wedge \kappa = p^{2}\int_{\mathbb{R}^{1,1}}\phi^{2}
\end{equation}
Likewise, under the topological twisting, three scalar fields become one-forms on \(M\).  Again, we can split the one-form bundle, and write the scalar part as \(\sigma\).  Then the kinetic term includes,
\begin{equation}
\int_{\mathbb{R}^{1,1}}\varphi^{2}\int_{M} d\kappa\wedge \kappa = p^{2}\int_{\mathbb{R}^{1,1}}\varphi^{2}
\end{equation}
Thus we end up with quadratic mass terms for both \(\phi\) and \(\varphi\).  But since these scalars are by definition the real and imaginary parts of \(\sigma\), this is precisely the same as the potential generated by the twisted superpotential in equation \ref{eqn:sigmapot}. Thus we have given an alternate way to understand the appearance of the quadratic terms in the twisted superpotential that break \(\mathcal{N}=(4,4)\) to \(\mathcal{N}=(2,2)\). 

\subsection{Computing the refined index}

The basic building block for the derivation of refined Chern-Simons theory in \cite{CSRefined}, is the value of the index on $M=R^2\times S^1$.  We will now explain how to compute it, in an arbitrary ADE case.

Note that with a flat metric on \(\mathbb{R}^{2}\) this geometry actually preserves even more supersymmetry, but we will allow an arbitrary metric on \(\mathbb{R}^{2}\) so that topological twisting is necessary and only \(\mathcal{N}=4\) supersymmetry is preserved.  Recall that the refined index is given by,
\begin{equation}
Z_{ref\;open}(X;q,t) = \textrm{Tr}\; (-1)^{2S_{1}}q^{S_{1}-S_{R}}t^{S_{R}-S_{2}}e^{-\beta H}
\end{equation}
and the contribution of an \(\mathcal{N}=2\) BPS state with charges \((S_{1}, S_{2}, S_{R})\) is given by a quantum dilogarithm,
\begin{equation}
\exp{\Big(\sum_{n=1}^{\infty}\frac{1}{n}\frac{(-1)^{2S_{1}}q^{n(S_{1}-S_{R})}t^{n(S_{R}-S_{2})}}{1-q^{n}}e^{nx}\Big)}
\end{equation}
where \(x\) is the complexified mass of the BPS state.  Now we can consider moving onto the Coulomb branch of the five-dimensional Yang-Mills theory; in the M-theory picture, this corresponds to separating the branes in the fiber direction over the \(S^{1}\).  This geometric deformation will become complexified by the Wilson line around the \(S^{1}\), so that altogether we have a complex scalar deformation.

On the Coulomb Branch, after performing the partial topological twist, there will be a massive three-dimensional \(\mathcal{N}=4\) BPS state for every \(W\)-boson in the Yang-Mills theory.  This \(\mathcal{N}=4\) state will decompose into two \(\mathcal{N}=2\) BPS states.  These two states are related by acting with supercharges in the \(\mathcal{N}=4\) algebra, but not in the \(\mathcal{N}=2\) algebra.  But from the discussion above, these supercharges will raise or lower the charge \(S_{R}-S_{2}\).  Therefore, altogether we have two BPS states of charges \((0,0,0)\) and \((+\frac{1}{2},-\frac{1}{2},+\frac{1}{2})\) under the \(SO(2)_{S_{1}}\times SO(2)_{S_{2}} \times SO(2)_{R}\) symmetry, up to an overall ambiguous shift in the ground state charge.  Therefore, the total partition function for the type-\(G\) \((2,0)\) theory is given by,
\begin{equation}
Z(\mathbb{R}^{2} \times S^{1}; G, x, q,t) = \exp{\Big(\sum_{\alpha>0}\sum_{n=1}^{\infty}\frac{1}{n}\frac{1-t^{n}}{1-q^{n}}e^{n\langle \alpha, x\rangle}}\Big)
\end{equation}
where the \(x\) variables are the complex Coulomb-branch parameters, and \(\alpha\) are the roots of the algebra \(G\).

We can also use this simple geometry to make another connection with the field theory interpretation of the refined Chern-Simons theory.  It is interesting to consider the \(\epsilon_{1} \to 0\), (\(q \to 1\)) limit.
In this limit the \(\Omega\)-deformation along the M5-brane is turned off, but it remains nonzero along the noncompact directions transverse to the brane.  For general three-manifolds, the free energy of the partition function then simply computes the twisted effective superpotential, as in the unrefined case \cite{Ooguri:1999bv}.  The only difference is that here, the mass of the BPS states contributing to \(\mathcal{W}_{eff}\) also depends on the R-charge because of the remaining \(\Omega\)-deformation in the transverse directions.  In fact, this corresponds to turning on a twisted mass equal to \(\epsilon_{2}(S_{R}-S_{2})\) for each field in the \(\mathcal{N}=2\) theory.  In our simple geometry, \(\mathbb{R}^{2} \times S^{1}\), taking this limit gives,
\begin{eqnarray}
\lim_{\epsilon_{1}\to 0}Z_{CS}(\mathbb{R}^{2} \times S^{1},G, x, q,t) & = & \mathrm{exp}\Bigg(-\frac{1}{\epsilon_{1}}\sum_{\alpha>0}\sum_{n=1}^{\infty}\frac{1}{n^{2}}\Big(e^{n\langle \alpha, x\rangle}-e^{n(\langle \alpha, x \rangle+\epsilon_{2})}\Big)\Bigg)
\end{eqnarray}
But these are precisely the contributions to the twisted effective superpotential from a vector and chiral multiplet in three dimensions \cite{Ooguri:1999bv, Nekrasov:2009uh}, if we identify the \(x_{I}\) as the twisted chiral superfield strengths.

Altogether, we have derived the contribution to the index for the solid torus geometry, which can be identified with \(\mathbb{R}^{2} \times S^{1}\), and we have also understood the effect of fibering the \(S^{1}\) nontrivially.  Putting this information together, we can glue together two solid tori to give the \(S^{3}\) geometry.  The gluing process leads to the refined Chern-Simons matrix model that was the basis for deriving more general amplitudes in \cite{CSRefined}.

\section{Macdonald polynomials \label{app:mac}}
Here we review some useful properties of the Macdonald polynomials associated to any semisimple lie algebra, \(\mathfrak{g}\) \cite{Macdonald1}.  For more details see \cite{macdonaldortho, KirillovNotes}.

We will denote the rank of \(\mathfrak{g}\) by \(r\), the root system of \(\mathfrak{g}\) by \(R\) and its positive part by \(R_{+}\).  Similarly, we denote the root lattice by \(Q\) and its positive part by \(Q_{+}\), and the weight lattice by \(P\) and its positive part by \(P_{+}\).  The root system lives in an \(r\)-dimensional vector space, \(V=\mathbb{R}^{r}\) and we denote the standard basis of \(V\) by \(\{\epsilon_{i}\}\), with the normalization that the \(\{\epsilon_{i}\}\) are the smallest vectors, such that they all sit inside \(P\).

We introduce \(r\) variables, \(\{x_{1}, \cdots, x_{r}\}\).  Formally, we can relate the \(x_{i}\) to the basis of V by the relationship, \(x_{i} \sim e^{\epsilon_{i}}\).  Since the Weyl group, \(W\), has a well defined action on the weight lattice of \(\mathfrak{g}\), this allows us to define an action of the Weyl group on \(\mathbb{Z}[x_{i},x_{i}^{-1}]\). 

Then we say that a polynomial, \(f(x_{1},\cdots, x_{r})\), is symmetric if \(f\) is invariant under the action of the Weyl group on the \(\{x_{i}\}\).  Note that in the case of \(\mathfrak{g} = \mathfrak{su}(N)\), this reduces to the usual definition of a symmetric polynomial since the Weyl group of \(\mathfrak{su}(N)\) is simply the symmetric group.  

The simplest example of such a symmetric polynomial is given by the monomial symmetric polynomials, \(m_{\lambda}\) ,which are associated to a representation of \(\mathfrak{g}\) with highest weight \(\lambda = \sum_{i}\lambda_{i}\epsilon_{i}\).  We use the notation that \(x^{\lambda} := x_{1}^{\lambda_{1}}\cdots x_{r}^{\lambda_{r}}\).  Then \(m_{\lambda}\) is defined by,

\begin{equation}
m_{\lambda} = \sum_{w\in W}x^{w(\lambda)}
\end{equation}
Note that this polynomial is symmetric by construction because of the sum over the Weyl group.

A slightly more sophisticated example is given by the character of a representation, \(\chi_{\lambda}\).  Recall that from the Weyl character formula, the character of \(\lambda\) can be written as,
\begin{equation}
\chi_{\lambda} = \frac{\sum_{w \in W}\epsilon(w)x^{w(\lambda+\rho)}}{\sum_{w \in W}\epsilon(w)x^{w(\rho)}}
\end{equation}
Although \(\chi_{\lambda}\) is usually described as a character, it can also be uniquely defined in another way that is closer to the definition of Macdonald polynomials.  We start by defining an inner product on the space of symmetric polynomials given by,
\begin{equation}
\langle f, g\rangle =  \frac{1}{\vert W \vert} \int_{0}^{2\pi} d\phi_{1} \cdots \int_{0}^{2\pi}d\phi_{r}\Delta(e^{i\phi_{1}},\cdots, e^{i\phi_{r}})f(e^{i\phi})g(e^{-i\phi})
\end{equation}
where
\begin{equation}
\Delta(x_{1},\cdots, x_{r}) = \prod_{\alpha \in R}(1-x^{\alpha})
\end{equation}
and where we use the shorthand \(f(x)\) for \(f(x_{1}, \cdots , x_{r})\).

Then we can uniquely define the \(\chi_{\lambda}\) to be the symmetric functions obeying a condition on the leading behavior and an orthogonality condition,
\begin{eqnarray}
\chi_{\lambda} & = & m_{\lambda} + \sum_{\mu < \lambda} K_{\lambda \mu}m_{\mu} \label{eqn:chi} \\
\langle\chi_{\lambda},\chi_{\nu}\rangle & = & 0 \;\;\;\;\;\textrm{for}\;\; \lambda \neq \nu
\end{eqnarray}
where \(K_{\lambda \mu}\) are arbitrary coefficients in the decomposition of the character.  By \(\mu < \lambda\), we mean that \(\lambda - \mu \in Q_{+}\).  Thus the first condition only restricts the leading behavior of \(\chi_{\lambda}\).

Now we are ready to finally define the Macdonald Polynomials in a similar way.  The Macdonald polynomials are Laurent polynomials in the \(\{x_{i}\}\), but also depend rationally on two additional variables, \(q\) and \(t\).  First we define a new measure,
\begin{equation}
\Delta_{q,t} = \prod_{\alpha \in R}\frac{(x^{\alpha};q)_{\infty}}{(x^{\alpha}t; q)_{\infty}}
\end{equation}
where we have used the q-Pochhammer symbol,
\begin{equation}
(a; q)_{\infty} = \prod_{r=0}^{\infty}(1-aq^{r})
\end{equation}
Although Macdonald polynomials are defined for arbitrary values of \(q\) and \(t\), the formulas simplify in the case when \(t=q^{\beta}\), with \(\beta\) a positive integer,
\begin{equation}
\Delta_{q,q^{\beta}} = \prod_{\alpha \in R}\prod_{m=0}^{\beta-1}(1-q^{m}x^{\alpha})
\end{equation}
Using this measure we can define a new inner product on the space of symmetric polynomials,
\begin{equation}
\langle f, g \rangle_{q,t} = \frac{1}{\vert W \vert} \int_{0}^{2\pi} d\phi_{1} \cdots \int_{0}^{2\pi}d\phi_{r} \Delta_{q, t}(e^{i\phi_{1}},\cdots, e^{i\phi_{r}}) f(e^{i\phi})g(e^{-i\phi})
\end{equation}

Then the Macdonald polynomials, \(M_{\lambda}\) are uniquely defined by the same leading order condition as in \ref{eqn:chi}, and the orthogonality condition with respect to the new inner product, 
\begin{eqnarray}
M_{\lambda} & = & m_{\lambda} + \sum_{\mu < \lambda} u_{\lambda \mu}m_{\mu} \\
\langle M_{\lambda}, M_{\nu}\rangle_{q,t} & = & 0 \;\;\;\;\;\textrm{for}\;\; \lambda \neq \nu
\end{eqnarray}
Note that this definition does not explicitly specify the inner product of \(M_{\lambda}\) with itself.  For the case of \(t=q^{\beta}\) with \(\beta\) a positive integer, this inner product, which we sometimes refer to as the metric, \(g_{\lambda}\), is equal to,
\begin{equation}
\langle M_{\lambda}, M_{\lambda} \rangle = g_{\lambda} \equiv \prod_{\alpha \in R_{+}}\prod_{m=0}^{\beta - 1}\frac{1-t^{\langle \rho, \alpha^{\vee}\rangle}q^{\langle \lambda, \alpha^{\vee}\rangle + m}}{1-t^{\langle \rho, \alpha^{\vee}\rangle}q^{\langle \lambda, \alpha^{\vee}\rangle - m}}
\end{equation}
where \(\rho\) is the Weyl vector, \(\rho = \frac{1}{2}\sum_{\alpha>0}\alpha\).  It will also be useful for knot computations to have a combinatorial expression for the metric that holds for general \(q\) and \(t\).  For the \(SO(2N)\) case, such a formula is given in Appendix B.  There exist relatively efficient algorithms for computing Macdonald polynomials for general root systems, such as the determinantal formula of \cite{vandiejen}.

As in the case of characters, the product of two Macdonald polynomials can be decomposed into a sum of Macdonald polynomials, where the coefficients are known as \((q, t)\) Littlewood-Richardson coefficients,
\begin{equation}
M_{\lambda}M_{\nu} = \sum_{\gamma}N^{\gamma}_{\lambda \nu}M_{\gamma}
\end{equation}
In general the computation of the \(N^{\gamma}_{\lambda \nu}\) is difficult, but for specific choices of representations, the Pieri formula gives an explicit expression for \(N\) \cite{Lassalle2009}.  

A weight, \(\lambda\), is known as minuscule if \(\langle \lambda, \alpha\rangle = 0\) or \(1\), for all \(\alpha \in R_{+}\).  Note that the fundamental weights, \(\omega_{i}\), are not necessarily minuscule since they obey, \(\langle \omega_{i}, \alpha_{j}^{\vee} \rangle = \delta_{ij}\) only for the dual of the simple roots, \(\alpha_{i}\), and not necessarily for all positive roots.  In fact, for the cases of \(E_{8}, F_{4},\) and \(G_{2}\), it is known that no minuscule weights exist.

However, when a minuscule weight, \(\omega\), does exist, then its product with any other weight, \(\lambda\) is given by,
\begin{equation}
M_{\omega}M_{\lambda} = \sum_{\begin{subarray}{c}
\tau \in W(\omega) \\
\lambda + \tau \in P^{+}
\end{subarray}}
N^{\lambda+\tau}_{\omega, \lambda}M_{\lambda + \tau}
\end{equation}
where,
\begin{equation}
N^{\lambda+\tau}_{\omega, \lambda} = \prod_{\begin{subarray}{c}
\alpha \in R_{+} \\
\langle \tau, \alpha^{\vee} \rangle = -1
\end{subarray}} \frac{1-t^{-1}q^{\langle \lambda + \rho, \alpha^{\vee}\rangle}}{1-q^{\langle \lambda + \rho, \alpha^{\vee}\rangle}}\frac{1-t q^{\langle \lambda + \rho, \alpha^{\vee}\rangle - 1}}{1-q^{\langle \lambda + \rho, \alpha^{\vee}\rangle - 1}}
\end{equation}

\section{Facts about $SO(2N)$ \label{app:so2n}}
Here we collect some useful facts about the lie algebra \(\mathfrak{so}(2N)\) and its Macdonald polynomials.  For \(\mathfrak{so}(2N)=D_{N}\) recall that the positive roots are given by, 
\begin{equation}
\{e_{i}-e_{j}, e_{i}+e_{j}\}, \; i<j
\end{equation}
and the Weyl vector is given by,
\begin{equation}
\rho = \sum_{i=1}^{N}(N-i)e_{i}
\end{equation}
The highest root is equal to,
\begin{equation}
\theta = e_{1}+e_{2}
\end{equation}
Finally, the fundamental weights , \(\omega_{i}\), are,
\begin{eqnarray}
\{\omega_{i} & \equiv & e_{1}+\cdots+e_{i}\}, \; 1\leq i \leq N-2 \\
\omega_{N-1} & = & \frac{1}{2}\Big(e_{1}+\cdots + e_{N}\Big) \nonumber \\
\omega_{N} & = & \frac{1}{2}\Big(e_{1}+\cdots + e_{N-1}-e_{N}\Big) \nonumber
\end{eqnarray}
In this section, we will write the weight of a generic representation as,
\begin{equation}
\lambda = \sum_{i=1}^{N}\lambda_{i}e_{i} = \sum_{i=1}^{N}\gamma_{i}\omega_{i}
\end{equation}
This means that the representations of \(D_{N}\) are indexed by \(\{\lambda_{i}\}\) which are either all integers or all half-integers, and satisfy,
\begin{equation}
\lambda_{1} \geq \lambda_{2} \geq \cdots \geq \lambda_{N-1} \geq \vert \lambda_{N} \vert
\end{equation}
In ordinary Chern-Simons theory, the Hilbert space of the theory on \(T^{2}\) is spanned by the integrable representations at level \(k\).  This result also holds true for the refined Chern-Simons theory, so for computations it is useful to know precisely which representations are integrable.  Recall that for a representation to be integrable, we require that,
\begin{equation}
(\lambda,\theta)\leq k
\end{equation}
where \(\lambda\) is the highest weight for the integrable representation, and \(\theta\) is the highest root.  Thus for \(D_{N}\), the integrability condition becomes,
\begin{equation}
\gamma_{1}+2\Big(\gamma_{2} + \cdots +\gamma_{N-2}\Big)+\gamma_{N-1}+\gamma_{N}\leq k
\end{equation}

For computations in the refined Chern-Simons theory, the Macdonald polynomials of \(D_{N}\) also play an important role.  The framework outlined in Appendix A is generally sufficient, but it is useful to have an explicit combinatorial formula for the metric even when \(\beta\) is not an integer in \(t=q^{\beta}\).  

Note from the structure of the fundamental weights, either all of the \(\{\lambda_{i}\}\) are integers or none of them are integers.  Also note that \(\lambda_{N}\) may be negative. If we define, \(\widetilde{\lambda}_{i} = \vert \lambda_{i} \vert\), then the \(\widetilde{\lambda}_{i}\) can naturally be described as a Young tableau when all the \(\lambda_{i}\) are integers.  The metric can then be expressed as a sum over the boxes of this tableau,
\begin{eqnarray}
\langle M_{\lambda}, M_{\lambda} \rangle = g_{\lambda} & = & \prod_{(i,j)\in \widetilde{\lambda}}\frac{(1- t^{2N-2i}q^{2j-1})(1-t^{N-i+1}q^{j-1})(1+t^{N-i}q^{j-1})(1-t^{N-i-1}q^{j})}{(1-t^{2N-2i-1}q^{2j})(1-t^{2N-2i-1}q^{2j-1})(1-t^{N-i}q^{j})} \nonumber \\
& & \cdot \frac{1-t^{\widetilde{\lambda}^{T}_{j}-i}q^{\widetilde{\lambda}_{i}-j+1}}{1-t^{\widetilde{\lambda}^{T}_{j}-i+1}q^{\widetilde{\lambda}_{i}-j}}\frac{1-t^{2N-\widetilde{\lambda}^{T}_{j}-i-1}q^{\widetilde{\lambda}_{i}+j}}{1-t^{2N-\widetilde{\lambda}^{T}_{j}-i}q^{\widetilde{\lambda}_{i}+j-1}}
\end{eqnarray}
Here, \(i\) is the vertical coordinate of the tableau and runs from \(1\) to \(N\), while \(j\) is the horizontal coordinate.  \(\widetilde{\lambda}_{i}\) is the length of the row \(i\), while  \(\widetilde{\lambda}^{T}_{j}\) denotes the number of boxes in column \(j\).

For the case when all of the \(\lambda_{i}\) are half-integers, we define, \(\widetilde{\lambda}_{i} = \vert \lambda_{i} \vert -1/2\), and again \(\widetilde{\lambda}\) can be interpreted as a Young tableau.  Then the corresponding formula for the metric is given by,
\begin{eqnarray}
\langle M_{\lambda}, M_{\lambda} \rangle = g_{\lambda} & = & C\prod_{(i,j)\in \widetilde{\lambda}}
\frac{1- t^{2N-2i}q^{2j}}{1-t^{2N-2i-1}q^{2j+1}}\frac{1-t^{2N-2i}q^{2j-1}}{1-t^{2N-2i-1}q^{2j}}\frac{1-t^{N-i+1}q^{j-1}}{1-t^{N-i}q^{j}}\frac{1-t^{N-i-1}q^{j+1}}{1-t^{N-i}q^{j}} \nonumber \\
& & \cdot \frac{1-t^{\widetilde{\lambda}^{T}_{j}-i}q^{\widetilde{\lambda}_{i}-j+1}}{1-t^{\widetilde{\lambda}^{T}_{j}-i+1}q^{\widetilde{\lambda}_{i}-j}}\frac{1-t^{2N-\widetilde{\lambda}^{T}_{j}-i-1}q^{\widetilde{\lambda}_{i}+j+1}}{1-t^{2N-\widetilde{\lambda}^{T}_{j}-i}q^{\widetilde{\lambda}_{i}+j}}
\end{eqnarray}
where \(C\) is an additional factor given by, 
\begin{equation}
C = \prod_{k=1}^{\lfloor N/2 \rfloor} \frac{1-t^{2k-2}q}{1-t^{2k-1}}\frac{1-t^{2k-2+2\lceil N/2 \rceil}}{1-t^{2k-3+2 \lceil N/2 \rceil}q}
\end{equation}

\section{Refined Indices \label{sec:index}}
Here we give details on the precise three- and five-dimensional indices that compute refined topological string partition functions.  In five dimensions this is simply the index studied by Nekrasov in \cite{Nekrasov:2002qd}.  Both indices are analogues of the four-dimensional protected spin character studied in \cite{Gaiotto:2010be}.

\subsection{Five Dimensional Indices}
Recall that the five-dimensional \(\mathcal{N}=1\) supersymmetry algebra consists of eight supercharges and includes an \(Sp(1)_{r}=SU(2)_{r}\) R-symmetry.  Upon dimensional reduction, the algebra is equivalent to four-dimensional \(\mathcal{N}=2\) supersymmetry.  For our purposes, it will be useful to rewrite the algebra in terms of four-dimensional notation, so that the supercharges can be organized as,
\begin{equation}
Q^{I}_{\alpha}, Q^{I}_{\dot{\alpha}}
\end{equation}
where \(I=1,2\) is the \(SU(2)_{R}\) index, while \(\alpha\) and \(\dot{\alpha}\) are the \(SO(4)=SU(2)_{l}\times SU(2)_{r}\) indices under rotations in four dimensions.  Then the supersymmetry algebra is given by,
\begin{eqnarray}
\{Q_{\alpha}^{I},\overline{Q}_{\dot{\beta}}^{J}\} & = & 2\delta^{IJ}\sigma_{\alpha\dot{\beta}}^{\mu}P_{\mu} \\
\{Q_{\alpha}^{I}, Q_{\beta}^{J}\} & = & 2\epsilon_{\alpha\beta}\epsilon^{IJ}(Z-iP_{5})
\end{eqnarray}
where \(Z\) is the \emph{real} five dimensional central charge.  Now we can study massive representations of this algebra, whose little group is \(SO(4)=SU(2)_{l} \times SU(2)_{r}\).  Then a generic long multiplet (with \(M \geq \vert Z \vert\)) transforms under \(SU(2)_{l} \times SU(2)_{r} \times SU(2)_{R}\) as,
\begin{equation}
(J_{l},J_{r};I_{R}) \otimes\Big((0,0;0)\oplus (0,0;1) \oplus (0,\frac{1}{2};\frac{1}{2}) \oplus (\frac{1}{2}, 0; \frac{1}{2}) \oplus (\frac{1}{2},\frac{1}{2};0)\Big)
\end{equation}
where \((J_{l},J_{r};I_{R})\) is an arbitrary representation.  In contrast, short left-handed BPS multiplets (with \(M = Z\)) take the form,
\begin{equation}
(J_{l},J_{r};I_{R})\otimes \Big((0,0;\frac{1}{2})\oplus (\frac{1}{2},0;0)\Big)
\end{equation}
while those with \(M=-Z\) will have the same structure but with the chirality flipped.

The unrefined index, which is related to the ordinary topological string, is given by
\begin{equation}
\mathrm{Tr}(-1)^{2(j_{l}+j_{r})}q^{2j_{l}}e^{-\beta H}
\end{equation}
The contribution of a long multiplet to this index is \(0\). In addition, a short right-handed multiplet contributes 0, while the fundamental left-handed multiplet contributes 
\begin{equation}
-\big(q^{1/2}-q^{-1/2}\big)^{2}e^{-\beta M}
\end{equation}
Thus, this gives a good index, since it only receives contributions from left-handed BPS states, while long multiplets cancel out of the trace. This is precisely the five-dimensional index in M-theory compactified on a Calabi-Yau that computes the ordinary topological string.

In order to extend this to a more refined index, we must use the R-symmetry, as in \cite{Nekrasov:2002qd}, which gives the index,
\begin{equation}
\mathrm{Tr}(-1)^{2(j_{l}+j_{r})}q_{1}^{2j_{l}}q_{2}^{2(j_{r}-S_{R})}e^{-\beta H}
\end{equation}
Again, we find that the long multiplets and the right-handed multiplets do not contribute to the index, while the fundamental left-handed multiplet now contributes,
\begin{equation}
(q_{2}+q_{2}^{-1}-q_{1}-q_{1}^{-1})e^{-\beta M}
\end{equation}
By using the definitions \(j_{l}=\frac{1}{2}(S_{1}-S_{2})\), \(j_{r}= \frac{1}{2}(S_{1}+S_{2})\), and taking \(q_{1}=\sqrt{qt}\), \(q_{2}=\sqrt{q/t}\), we can rewrite this in the form more natural for refined topological string theory,
\begin{equation}
\mathrm{Tr} (-1)^{F}q^{S_{1}-S_{R}}t^{S_{R}-S_{2}}e^{-\beta H}
\end{equation}
and the contribution from a BPS multiplet becomes, 
\begin{equation}
-\Big(\sqrt{q}-\frac{1}{\sqrt{q}}\Big)\Big(\sqrt{t}-\frac{1}{\sqrt{t}}\Big)e^{-\beta H}
\end{equation}

\subsection{Three Dimensional Indices}
Here we collect the details on indices for three dimensional \(\mathcal{N}=2\) supersymmetry, since this is the case of primary interest for refined Chern-Simons theory.  The supersymmetry algebra is given by dimensionally reducing \(\mathcal{N}=1\) supersymmetry in four dimensions,
\begin{eqnarray}
\{Q_{\alpha},\overline{Q}_{\beta}\} & = & 2\sigma^{\mu}_{\alpha\beta}P_{\mu} + 2i\epsilon_{\alpha\beta}Z \\
\{Q_{\alpha},Q_{\beta}\} & = & 0 \\
\{\overline{Q}_{\alpha}, \overline{Q}_{\beta}\} & = & 0
\end{eqnarray}
The \(Q\) are complex spinors with charge \(\pm \frac{1}{2}\) under the two-dimensional rotation group, \(U(1)_{S}\), and as in four dimensions there is a \(U(1)_{R}\) symmetry that rotates \(Q_{\alpha}\) and \(\overline{Q}_{\alpha}\) \footnote{In this appendix, we use notations that are more appropriate for the three-dimensional perspective, instead of the natural M-theory notations used in the body of the paper.  The relationship is \((S, R, r) \leftrightarrow (S_{1}, S_{2}, S_{R})\).}.  Then the simplest massive long (\(M>\vert Z \vert\)) representation transforms under \(U(1)_{S} \times U(1)_{R}\) as \((0;\pm\frac{1}{2})\oplus(\pm \frac{1}{2};0)\).  As usual, a generic long representation is given by tensoring this with an arbitrary representation of \(U(1)_{S} \times U(1)_{R}\),
\begin{equation}
(S;R) \otimes \Big((0;\pm\frac{1}{2})\oplus(\pm \frac{1}{2};0)\Big)
\end{equation}
There are also short BPS multiplets, each containing one bosonic and one fermionic degree of freedom.  The right short representations are given by \((S;R)\otimes\big((0;-1/2)\oplus(+1/2;0)\big)\) and the left short representations are given by \((S;R)\otimes\big((0;-1/2)\oplus(-1/2;0)\big)\).

Now consider the unrefined index,
\begin{equation}
\textrm{Tr}(-1)^{2S}q^{S-R}e^{-\beta H}
\end{equation}

It can be seen that the long representation makes a contribution of 0 to this index, as do the right short multiplets, but the simplest left short representation contributes \((q^{1/2}-q^{-1/2})e^{-\beta M}\).  This index is related to the open topological string, and thus, to unrefined Chern-Simons theory.

Generically, this unrefined index is the best that we can do.  However, as explained in Appendix \ref{sec:twist}, an additional symmetry, \(U(1)_{r}\), appears upon compactifying the \((2, 0)\) theory on a Seifert Manifold.  In this case, the supercharges have the quantum numbers shown in Table \ref{superchargetab}.

\begin{table}[htp]
\centering
\begin{tabular}[c]{| c | | c | c | c |}
\hline
& $2S$ & $2R$ & $2r$ \\
\hline
$Q^{1}_{+}$ & $+1$ & $+1$ & $+1$ \\
\hline
$Q^{1}_{-}$ & $-1$ & $+1$ & $+1$ \\
\hline
$Q^{2}_{+}$ & $+1$ & $-1$ & $-1$ \\
\hline
$Q^{2}_{-}$ & $-1$ & $-1$ & $-1$ \\
\hline
\end{tabular}
\caption{Quantum Numbers for Three-Dimensional $\mathcal{N}=2$ Supersymmetry \label{superchargetab}}
\end{table}
We can now form a refined index
\begin{equation}
\textrm{Tr}(-1)^{2S}q^{S-r}t^{r-R}
\end{equation}
Note that \((r-R)\) is a flavor symmetry, since none of the supercharges are charged under it.  Then it is straightforward to show that long multiplets and right short multiplets make no contribution to this improved index.  The only contribution comes from the fundamental left short multiplet, which contributes \((q^{1/2}-q^{-1/2})\) as before.  

However, there will generically still be t-dependence, since the BPS states may be charged under the new flavor symmetry.  It is precisely this index that computes the refined Chern-Simons theory.

\bibliography{RefinedOrientifolds}{}

\end{document}